\documentclass{pasj00}

\begin{document}
\SetRunningHead{T. Nomura, J. Naito \& H. Shibahashi}
{Numerical simulations of line-profile variation in roAp stars}
\Accepted{2011/08/25}

\title{Numerical simulations of line-profile variation\\ 
beyond a single-surface approximation for oscillations in roAp stars}

\author{Takashi \textsc{Nomura}, Jun \textsc{Naito}, Hiromoto \textsc{Shibahashi}}
\affil{Department of Astronomy, The University of Tokyo, Tokyo 113-0033, Japan}
\email{shibahashi@astron.s.u-tokyo.ac.jp}


\maketitle

\begin{abstract}
\noindent 
Prior to the last decade, most observations of roAp stars have concerned the light variations. Recently some new, striking results of spectroscopic observations with high time resolution, high spectral dispersion, and a high signal-to-noise ratio became available. Since the oscillations found in roAp stars are high overtones, the vertical wavelengths of the oscillations are so short that the amplitude and phase of the variation of each spectroscopic line are highly dependent on the level of the line profile. Hence, analyses of the variation of the spectroscopic lines of roAp stars potentially provide us with new information about the vertical structure of the atmosphere of these stars. In order to extract such information, a numerical simulation of the line-profile variation beyond a single-surface approximation is necessary. We carried out a numerical simulation of line-profile variation by taking account of the finite thickness of the line forming layer. We demonstrated how effective this treatment is, by comparing the simulation with the observed line profiles. 
\end{abstract}

\section{Introduction}
\label{sec:1}
\noindent
A fraction of A-type stars, called Ap stars with the subscript ``p'', indicating ``peculiar'', show strong absorption lines of unusual elements. Spectral analyses of these stars indicate that rare earth elements are very abundant, at least at a certain atmospheric level.  Such anomalous spectra seen in Ap stars are thought to be a consequence of the levitation and resultant accumulation of those elements into the high atmosphere due to radiation pressure. The photon momentum is selectively transferred to the ions of those elements through their appropriate absorption lines. Many of the Ap stars have global magnetic fields, of which the apparent strength is on the order of 10\,mT (=\,1\,kG), and varies periodically and mostly sinusoidally on a time scale of days to decades. This periodic variation is likely to be caused by rotation of the star in which the magnetic axis is inclined to the rotational axis. The spectrum and mean brightness vary synchronously with the magnetic field strength. Such synchronous variation is thought to be a consequence of the fact that the atmospheres of magnetic Ap stars are laterally inhomogeneous, so that some chemical elements are selectively levitated near the magnetic polar region, while others are levitated near the magnetic equatorial zone, because the presence of magnetic fields strongly influence diffusion. This phenomenological model is known as the ``oblique rotator model'', and it is now widely accepted as explanating the magnetic, spectral and the mean light variations seen in Ap stars.  

A-type stars occupy the region of the Hertzsprung-Russell (HR) diagram where the Cepheid instability strip crosses the zero-age main sequence. The A-type stars located in the instability strip are thus expected to pulsate, and there are indeed many pulsating stars -- the $\delta$ Scuti type variables -- whose pulsation period is on the order of two hours. Pulsational variability has long been sought in Ap stars. It is almost thirty years ago since rapid light variations in the magnetic holmium star HD\,101065 were discovered by \citet{Kurtz78}. Similar rapid light variations with periods in the range of 6 to 12 min were discovered soon thereafter in several cool magnetic Ap stars (\cite{Kurtz82}), and the concept of ``rapidly oscillating Ap stars'' was established.  Currently, 40 Ap  stars are known to pulsate. The period range of the oscillations found in Ap stars is 6--21 min and is substantially shorter than the dynamical timescale, $2 \pi [R^3/(GM)]^{1/2}$, which for main-sequence A-type stars is typically about two hours. This is obviously the reason why the stars are called rapidly oscillating Ap stars (\cite{Kurtz82}), or ``roAp stars'' for short (\cite{MKW87}).  This means that the oscillations in Ap stars are of very high overtones.

Another noticeable characteristic of oscillations in Ap stars is that, in many cases, the pulsation amplitudes of light variations vary synchronously with the apparent magnetic field strength of the star.  In some cases, even though no magnetic field has yet been detected, the pulsation amplitudes are found to vary periodically with a time scale of several days. \citet{Kurtz82} showed that amplitude modulation of the rapid oscillations in Ap stars is well-explained in terms of an axisymmetric dipole mode whose axis is aligned with the magnetic axis and inclined to the rotational axis of the star. It is then obvious that the rapid oscillations in Ap stars are strongly influenced by stellar magnetism; in this sense, the roAp stars are unique among the various types of pulsating star. 
Recent long time-series photometry has enabled us to perform improved asteroseismology of this class of stars (\cite{Kurtz_etal05b}; \cite{Cameron_etal06}; \cite{Gruberbauer_etal08}; \cite{Huber_etal08}; {\cite{Balona_etal11a}, \yearcite{Balona_etal11b}; \cite{Kurtz_etal11}).

As mentioned above, the atmospheres of Ap stars are laterally chemically inhomogeneous, offering the potential to map the oscillations. Furthermore, since the oscillations found in roAp stars are high overtones, the vertical wavelengths of the oscillations are so short that the amplitude and phase of the oscillations are highly dependent on the atmospheric level. Hence, the analyses of rapid oscillations in roAp stars potentially provide us with a new tool to diagnose the inhomogeneous ---both laterally and vertically--- atmospheres of Ap stars, which are governed by radiative levitation and strong magnetic fields. High-resolution spectroscopic observations are required for this purpose. 

Prior to the last decade, most observations of roAp stars have concerned the light variations [see reviews; e.g., \citet{Kurtz_Martinez00}]. During the early phase of spectroscopic investigations, observations were done with low-resolution spectroscopy. In the case of $\alpha$\,Cir, it was found that some lines are apparently pulsating in anti-phase with others. \citet{Baldry_etal98} interpreted this in terms of a high-overtone standing wave with a velocity node in the atmosphere of the star.  Bisector measurements of the H\,$\alpha$ line show that the velocity amplitude and the phase of the principal oscillation mode vary significantly, depending on height in the H\,$\alpha$ line.  This fact confirms the existence of a radial node in the atmosphere of the star.  Therefore, it has been suggested that the use of spectral lines that form at different levels in the atmosphere can be used to gain depth information about the eigenfunction.

Thanks to big telescopes and high resolution spectrographs, recently some new, striking results of spectroscopic observations of oscillations of roAp stars with high time resolution, high spectral dispersion, and a high signal-to-noise ratio became available
(\cite{Savanov_etal99}; \cite{Kochukhov_Ryabchikova01a}, 
\yearcite{Kochukhov_Ryabchikova01b}; \cite{Balona_Zima02}; 
\cite{Kochukhov_etal02}; \cite{Balona02}; \cite{Balona_Laney03}; 
\cite{Kurtz_etal03}, \yearcite{Kurtz_etal05a}, \yearcite{Kurtz_etal05c}; \cite{Mkrtichian_etal03}; 
\cite{Sachkov_etal04}; \cite{Elkin_etal05a}; \cite{Elkin_etal05b}; 
\cite{Hatzes_Mkrtichian05}; \cite{Mkrtichian_Hatzes05}; 
\cite{Kurtz_etal06};
\cite{Kochukhov_etal2007};
\cite{Kurtz_etal07};
\cite{Ryabchikova_etal07a}, \yearcite{Ryabchikova_etal07b}; \cite{Elkin_etal08}; \cite{Mkrtichian_etal08}; \cite{Sachkov_etal08}; \cite{Freyhammer_etal08}, \yearcite{Freyhammer_etal09};  \cite{Kochukhov_etal2008}, \yearcite{Kochukhov_etal2009}).
These time-series spectroscopy observations clearly show that the amplitude and phase of variation of each spectroscopic line are indeed dependent on the intensity level of the line profile. Hence, the variation of the spectroscopic lines of roAp stars is a very promising information source about the vertical structure of the atmosphere of these stars as well as the magneto-acoustic oscillations in these stars. 

Modeling of line-profile variations (LPV) has so far been developed mainly to describe the observed variations among O, B variables,  ---$\beta$ Cephei stars, SPB stars, $\zeta$ Ophiuchi variables, which are nonradial pulsators oscillating with low order p-, or g-modes (\cite{Osaki71}; \cite{Stamford_Watson76}, \yearcite{Stamford_Watson77}; \cite{Smith77}; \cite{Kubiak78}; \cite{Vogt_Penrod83}; \cite{Balona86a}, \yearcite{Balona86b}, \yearcite{Balona87}; \cite{Kambe_Osaki88}; \cite{Aerts_Waelkens93}; \cite{Schrijvers_etal97}; \cite{Townsend97}). In these cases, contrary to the case of roAp stars, the vertical wavelengths of the oscillations are much longer than the thickness of the line-forming layer of each spectral line, and hence the line-forming layer at a given longitude and latitude on the stellar surface moves up and down uniformly throughout the layer. In order to see the influence of such oscillations on the spectral line profiles, we  do not  have to model the line shape in detail. We may ignore the thickness of the line-forming layer, and may treat the layer as a single surface to take account of the motion in the Doppler shift of the spectral line. Actually, in the standard procedure of line-profile modeling of these stars, it is sufficient to approximate the intrinsic line shape as a Gaussian profile to see the Doppler effect of oscillations. Beside that, the symmetry axis of their oscillations coincides with the rotation axis of the star, and hence the aspect angle of the oscillation axis does not change at all with time. 

However, the situation is different in the case of roAp stars, of which oscillations have vertical wavelengths as short as the thickness of the line-forming layers. In order to extract information about the structure of the atmosphere from spectroscopic observations of oscillations with high time resolution, a numerical simulation of line-profile variation beyond a single surface approximation is necessary. Furthermore, the aspect angle of the symmetry axis of the oscillation patterns changes with time along with rotation of the star. Therefore, we need to model properly the line profiles of roAp stars to extract asteroseismic information from the spectroscopic analyses of the time series of line-profile variation of these stars (\cite{Shibahashi08}). 

The main purpose of the present work is  to carry out numerical simulations of line-profile variation by taking account of the finite thickness of the line-forming layer. We demonstrated how effective this treatment is, by comparing the simulation with the observed line profiles. 
The procedure of calculating the intrinsic line profiles is outlined in section\,\ref{sec:2}. The equation of radiative transfer was numerically solved. The recipes for computation of the absorption coefficients are described in section\,\ref{sec:3}. We will verify the numerical code by comparing the computed line profile with observations in section\,\ref{sec:4}, and demonstrate the difference in LPV caused by motion with different vertical wavelengths. More detailed calculations are described in section\,\ref{sec:5} based on a realistic model of the A-type star. The variation in line bisectors is discussed in section\,\ref{sec:6}, and a comparison of the simulation results with observations is made. Future prospects are discussed in section\,\ref{sec:7}.

\section{Procedure of calculation of intrinsic line profile}
\label{sec:2}
\subsection{Outline}
\label{sec:2.1}
\noindent
It may be instructive to outline here a procedure to calculate the intrinsic line profile. We divide the stellar disk image by a mesh into small square cells, which are labelled by the $(x, y)$-coordinates. The energy flux going from each cell toward the observer per second in a unit wavelength at the wavelength $\lambda$ is calculated from specific intensity of radiation, $I_\lambda$, which is described by the equation of radiative transfer,
\begin{equation}
{{\partial I_\lambda (x, y, z)}\over{\partial z}}
=
-(\kappa_\lambda+\ell_\lambda+\sigma_\lambda)\rho I_\lambda
+j_\lambda \rho,
\label{eq:1}
\end{equation} 
where $z$ denotes the path length, and $\rho$ is the density in mass per unit volume; $\kappa_\lambda$ and $\ell_\lambda$ mean the continuous absorption coefficient per mass and the line absorption coefficient per mass, respectively, $\sigma_\lambda$ denotes the scattering coefficient, and $j_\lambda$ is the total emission coefficient. Since the beam path of light is different from cell to cell, $I_\lambda$ is regarded as being a function of $(x,y)$ and the derivative is taken as a partial one in equation (\ref{eq:1}) (see figure\,\ref{fig:1}). By dividing both sides of (\ref{eq:1}) by $-(\kappa_\lambda+\ell_\lambda+\sigma_\lambda)\rho$, we can rewrite the equation of radiative transfer in the form of
\begin{equation}
{{\partial I_\lambda (x, y, \tau_{\lambda z})}\over{\partial \tau_{\lambda z}}}
= I_\lambda - S_\lambda,
\label{eq:2}
\end{equation}
where $S_\lambda$ is the source function defined by
\begin{equation}
S_\lambda \equiv j_\lambda/(\kappa_\lambda+\ell_\lambda+\sigma_\lambda)
\label{eq:3}
\end{equation}
and 
\begin{equation}
d\tau_{\lambda z} = -(\kappa_\lambda+\ell_\lambda+\sigma_\lambda)\rho dz.
\label{eq:4}
\end{equation}
With $d\tau_{\lambda z}$, we introduce the optical depth, $\tau_{\lambda z}$, looking from the outside in, along the path for the outgoing beam of light. The intensity at the observer's site, $I_\lambda (x,y,0)$, is derived by an integral of equation (\ref{eq:2}) with respect to $\tau_{\lambda z}$ from $0$ to $\infty$, and is expressed as
\begin{equation}
I_\lambda (x,y,0) = \int_0^\infty S_\lambda (\tau_{\lambda z}) \exp(-\tau_{\lambda z}) d\tau_{\lambda z}.
\label{eq:5}
\end{equation}
The total energy flux going outward toward the observer per second in a unit wavelength band is then obtained by an integral of $I_\lambda (x,y,0)$ over the stellar disk,
\begin{eqnarray}
F_\lambda &\propto& \int_{\rm disk} I_\lambda(x,y,0) \,dx\,dy 
\nonumber \\
&=&
\int_0^\infty \int\int S_\lambda(\tau_{\lambda z}) \exp(-\tau_{\lambda z})
\,dx\,dy\,d\tau_{\lambda z}.
\label{eq:6}
\end{eqnarray} 
\begin{figure}[tb]
\begin{center}
\FigureFile(90mm,90mm){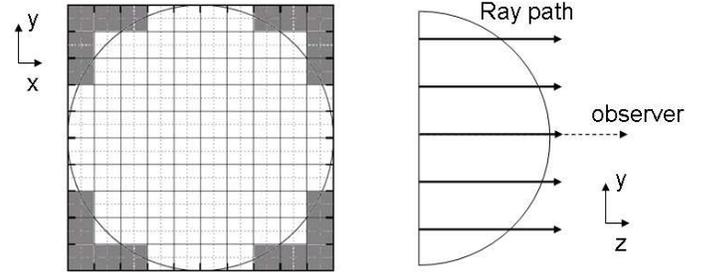}
\end{center}
\caption{Stellar disk image subdivided by a mesh coordinated with $(x,y)$ into small square cells. At each cell, the energy flux going toward the observer per second in a unit wavelength is calculated by integrating the specific intensity of radiation along the path in the direction of $z$-axis, which is toward the observer. }
\label{fig:1}
\end{figure}

In order to theoretically determine the profile of a spectral line for a given model atmosphere, we need to estimate the background, continuum, radiative flux from the star at the wavelength of the spectral line of interest and to estimate the flux at the same wavelength by taking account of both the line absorption and continuum absorption. The former is calculated by setting $\ell_\lambda=0$. Then, the line profile against the background continuum spectrum is given by residual flux, defined as the ratio of these two fluxes.

\subsection{Source function}
\label{sec:2.2}
\noindent
The total emission, $j_\lambda$, is composed of the thermal emission and the contribution of photons coming there by scattering from other directions. As for the former, we assume here for the sake of simplicity local thermodynamic equilibrium (LTE). Then, the thermal emission is described by the Planck function, $B_\lambda$. As for the latter, we assume coherent scattering, which does not change the wavelength of photons after scattering, and also isotropic scattering for simplicity. In these approximations, $j_\lambda$ is written as
\begin{equation}
j_\lambda=(\kappa_\lambda+\ell_\lambda)B_\lambda+\sigma_\lambda J_\lambda,
\label{eq:7}
\end{equation} 
where $J_\lambda$ is the mean intensity, defined by
\begin{equation}
J_\lambda \equiv \int I_\lambda {{d\Omega}\over{4\pi}}.
\label{eq:8}
\end{equation}
Since $J_\lambda$ becomes sufficiently close to the Planck function in the deep layers, we approximate $J_\lambda$ to be $B_\lambda$.
Then, total energy flux (\ref{eq:6}) is calculated by
\begin{eqnarray}
F_\lambda &\propto&
\int_0^\infty \int\int B_\lambda[T(x, y, z)]
\exp(-\tau_{\lambda z}) \,dx\,dy\,d\tau_{\lambda z}
\nonumber \\
&=&
\int_0^\infty \int\int B_\lambda[T(x, y, z)]
\exp(-\tau_{\lambda z}) 
\nonumber \\
& & \times
(\kappa_\lambda+\ell_\lambda+\sigma_\lambda)\rho\,dx\,dy\,dz.
\label{eq:9}
\end{eqnarray}
In the following numerical simulations, the integral over the disk, $dx\,dy$ in the above expression, is evaluated as summations with respect to the cells, and the integral over the beam path, $d\tau_{\lambda z}$, is the summation with respect to the thin sliced layers piled along the $z$-direction. In practice, since we are interested in low-degree p-modes with high order, the disk image is subdivided into $32\times 32$ square cells, and the volume is sliced by 200-300 thin zones in the $z$-direction. 

\subsection{Influence of motion}
\label{sec:2.3}
\noindent
Our main interest is distortion of the line profiles due to oscillations in the stellar atmosphere. Global motion in the stellar atmosphere leads to a Doppler shift for light emitted from there. Besides that, in the case of stellar pulsation, the temperature and any other thermodynamical quantities vary with the motion, and consequently the local surface brightness varies as well. Variations in the surface area and in the surface normal due to oscillation may also produce some variations of the line profile, but these effects other than the Doppler effect are probably minor. In this paper, we ignore them for the sake of simplicity, and restrict ourselves to an investigation of the Doppler effect on the line profiles. 
To take account of the Doppler shift corresponding to the line-of-sight velocity, we only have to shift the wavelength in equations (\ref{eq:4}) and (\ref{eq:9}) by the Doppler shift,
\begin{equation}
\lambda \longrightarrow \lambda \times [1-\mbox{\boldmath$v$}(x,y,z)\cdot\mbox{\boldmath${\rm e}$}_z]/c,
\label{eq:10}
\end{equation}
where $\mbox{\boldmath$v$}(x,y,z)$ denotes the velocity field at the coordinate $(x,y,z)$, $c$ is the speed of light and $\mbox{\boldmath${\rm e}$}_z$ is the unit vector toward the $z$-direction (see figure\,\ref{fig:1}).  
The velocity field, $\mbox{\boldmath$v$}$, is the sum of the velocity due to oscillation and stellar rotation. Convection and turbulence should also be taken into account, if any. However, in the case of A-type stars, surface convection or turbulence is not expected.

\section{Preparation: Absorption and scattering coefficients}
\label{sec:3}
\subsection{Continuum absorption coefficient}
\label{sec:3.1}
\noindent
We restrict ourselves to consider only the wavelength range of  visual light. In this range, for $T > 5000$\,K, the main sources of the continuum absorption coefficient are free-free and bound-free absorption of hydrogen atoms and of negative hydrogen ions. Following \citet{Gray05}, we calculate the absorption coefficient in units of area per neutral hydrogen atom for the free-free absorption and for the bound-free absorption, respectively, as follws:
\begin{equation}
\kappa ({\rm H}_{\rm ff}) = \alpha_0 \lambda^3 g_{\rm ff}
{{\log {\rm e}}\over{2\Theta I}}{10}^{-\Theta I},
\label{eq:11}
\end{equation}
\begin{equation}
\kappa ({\rm H}_{\rm bf}) = \alpha_0 \sum_{n=n_0}^{\infty} {{\lambda^3}\over{n^3}}
g_{\rm bf} {10}^{-\Theta \chi_n},
\label{eq:12}
\end{equation}
where $\Theta \equiv \log {\rm e}/kT = 5040/T$, $I$ is the ionization energy, in units of eV, of hydrogen (13.6\,eV), $\chi_n$ is the excitation potential, in units of eV, from the ground state $(n=1)$ to the $n$-th state [that is, $\chi_n = I(1-n^{-2})$)],  and $n_0$ denotes the smallest integer satisfying $\lambda < n_0^2/R_\infty$, where $R_\infty$ means the Rydberg constant. Here $g_{{\rm ff}}$ and $g_{{\rm bf}}$ are the Gaunt factors, given by
\begin{equation}
g_{\rm ff}=1+{{0.3456}\over{(\lambda R_\infty)^{1/3}}}
\left( {{\lambda kT}\over{hc}}+{{1}\over{2}}\right)
\label{eq:13}
\end{equation}
and
\begin{equation}
g_{\rm bf}=1-{{0.3456}\over{(\lambda R_\infty)^{1/3}}}
\left( {{\lambda R_\infty}\over{n^2}}-{{1}\over{2}} \right),
\label{eq:14}
\end{equation}
respectively \citep{Menzel_Pekeris35},
and $\alpha_0 \equiv 32\pi^2e^6R_\infty/[3^{3/2}(4\pi \varepsilon_0)^3 h^2c^3]$, where $e$ is the electron charge, $\varepsilon_0$ is the vacuum permittivity, $h$ is Planck's constant, and $k$ is Boltzmann's constant.

As for negative hydrogen, we adopt the following expression for the absorption coefficient in units of area (m$^2$) per neutral hydrogen atom for the bound-free absorption and for the free-free absorption \citep{Gray05}:
\begin{equation}
\kappa ({\rm H}_{{\rm bf}}^{-}) = 
4.158\times {10}^{-31} \alpha_{\rm bf} P_{\rm e} \Theta^{5/2} {10}^{0.754\Theta}  
\label{eq:15}
\end{equation}
and
\begin{equation}
\kappa ({\rm H}_{{\rm ff}}^{-}) =
{10}^{-29} P_{\rm e} {10}^{f_0+f_1\log\Theta +f_2\log^2\Theta}.  
\label{eq:16}
\end{equation}
Here, $P_{\rm e}$ is the electron pressure, in units of Pa. The wavelength dependence is given for each case in terms of a polynomial, which fits sufficiently well to the detailedly calculated absorption coefficients \citep{Gray05}; $\alpha_{\rm bf}$ in the case of bound-free absorption given by  
\begin{equation}
\alpha_{{\rm bf}} = \sum_{i=0}^6 a_i \lambda^i,
\label{eq:17}
\end{equation}
and $f_i$ $(i=0, 1, 2)$ in the case of free-free absorption given by
\begin{equation}
f_i=\sum_{j=0}^4 f_{i,j}\log^j\lambda,
\label{eq:18}
\end{equation}
where the wavelength, $\lambda$, is given in units of {\AA}.
The numerical constants of the coefficients $\{a_i\}$ and the fitting formulae for $f_0$, $f_1$, and $f_2$ are summarized in table\,\ref{tab:1} and in table\,\ref{tab:2}, respectively.

\begin{table}
\caption{Numerical values of the coefficients $\{a_i\}$ appearing in the fitting formula (\ref{eq:17}). 
Adopted from \citet{Gray05}.}
\label{tab:1}
\begin{center}
\begin{tabular}{cr@{.}l}
\hline
coefficient & \multicolumn{2}{c} {numerical value} \\ \hline
$a_0$ & $1$&$99654$ \\
$a_1$ & $-1$&$18267\times {10}^{-5}$ \\
$a_2$ & $2$&$64243\times  {10}^{-6}$ \\
$a_3$ & $-4$&$40524\times {10}^{-10}$ \\
$a_4$ & $3$&$23992\times {10}^{-14}$ \\
$a_5$ & $-1$&$39568\times {10}^{-14}$ \\
$a_6$ & $2$&$78701\times {10}^{-23}$ \\ \hline
\end{tabular}
\end{center}
\end{table}

\begin{table*}
\caption{Coefficients $f_{i,j}$ of the fitting formulae for $f_0$, $f_1$ and $f_2$ appearing in the formula (\ref{eq:18}). Adopted from \citet{Gray05}.}
\label{tab:2}
\begin{center}
\begin{tabular}{c r@{.}l r@{.}l r@{.}l r@{.}l c}
\hline
$j$  & \multicolumn{2}{c} {0} & \multicolumn{2}{c} {1} & \multicolumn{2}{c} {2} & \multicolumn{2}{c} {3} & 4  \\ \hline
$f_0$ & $-2$&$2763$ & $-1$&$6850$ & $0$&$76661$ & $-0$&$053346$ & -\\
$f_1$ & $15$&$2827$ &$ -9$&$2846$ & $1$&$99381$ & $-0$&$142631$ & -\\
$f_2$ & $-197$&$789$ & $190$&$266$ & $-67$&$9775$ & $10$&$6913$ & $-0.625151$ \\
\hline
\end{tabular}
\end{center}
\end{table*}

It should be noted here that the absorption processes are affected by stimulated emission and the absorption actually produced is lower by a factor of $\{1-\exp[-hc/(\lambda kT)]\}$. Also, it should be noted here that, following \citet{Gray05},  the absorption coefficients given in this subsection by equations (\ref{eq:11}), (\ref{eq:12}), (\ref{eq:15}) and (\ref{eq:16}) are those for per neutral hydrogen atom. Therefore, we have to convert them to those per mass, denoted by $\kappa_\lambda$, discussed in the previous section, by multiplying  a factor $n_{{\rm H}\emissiontype{I}}/\rho$, where $n_{{\rm H}\emissiontype{I}}$ denotes the number density of neutral hydrogen atoms.

\subsection{Line absorption coefficient}
\label{sec:3.2}
\subsubsection{General description}
\label{sec:3.2.1}
\noindent
The mass absorption coefficient, $\ell_{\lambda}$, expressed in area per unit mass, due to the  transition from the $l$-th excited level of the $k$-th ion of the element $j$, can be written with the help of the number density of the ions in the state of interest, $n_{jkl}$, as
\begin{eqnarray}
\rho \ell_{\lambda, jkl} &=& \alpha_\lambda n_{jkl}
\left\{1-\exp\left[-hc/(\lambda kT\right]\right\}
\nonumber \\
&=&
\alpha_\lambda n_j {{n_{jk}}\over{n_j}}{{n_{jkl}}\over{n_{jk}} }
\left\{ 1-\exp\left[-hc/(\lambda kT)\right]\right\},
\label{eq:19}
\end{eqnarray}
where $\alpha_\lambda$ denotes the absorption coefficient per atom, $\{1-\exp[-hc/(\lambda kT)]\}$ expresses the effect of stimulated emission, and $n_j$ and $n_{jk}$ are defined by
\begin{equation}
n_j \equiv \sum_k n_{jk}
\label{eq:20}
\end{equation}
and
\begin{equation}
n_{jk} \equiv \sum_l n_{jkl},
\label{eq:21}
\end{equation}
respectively.
Here, $n_{jk}/n_j$ represents the ionization degree of the $k$-th ion of the element $j$, and it is calculated with Saha's equation. The ratio $n_{jkl}/n_{jk}$ gives what fraction of the ions are at the $l$-th excited level among the $k$-th ion of the element $j$; it is given by 
\begin{equation}
{{n_{jkl}}\over{n_{jk}}}={{g_{jkl}}\over{U_{jk}(T)}}\exp\left(-{{\chi_{jkl}}\over{kT}}\right),
\label{eq:22}
\end{equation}
where $g_{jkl}$ denotes the statistical weight of the $l$-th level, $\chi_{jkl}$ is the excitation potential of the $l$-th level, and $U_{jk}(T)$ means the distribution function, defined by
\begin{equation}
U_{jk}(T)=\sum_l g_{jkl}\exp\left(-{{\chi_{jkl}}\over{kT}}\right).
\label{eq:23}
\end{equation}

\subsubsection{Line broadening}
\label{sec:3.2.2}
\noindent
Each of absorption lines is intrinsically  broadened for many reasons. One of the main causes is the fact that the lifetime of the concerning atomic energy level is not infinite due to the uncertainty principle, partly reflecting radiation damping, which is often called natural damping, due to the interaction of light with atoms, and partly reflecting collisions of atoms absorbing the light with other particles. Another main cause is the Doppler effect due to the thermal motion of atoms, of which the velocity distribution is the Maxwellian distribution. The effects of natural damping and of collisional damping lead to a Lorentzian profile, characterized by the frequency of the line center, $\nu_0$, and the damping parameter, $\Gamma$. 

\subsubsection{Voigt profile}
\label{sec:3.2.3}
\noindent
The combined effect of the natural damping and the thermal Doppler broadening is described by a convolution of the Lorentzian and the Maxwellian profiles. It is known that this convolution integral is reduced to the Hjerting function, $H(u,a)$, which is defined by
\begin{equation}
H(u,a)={{a}\over{\pi}}\int_{-\infty}^{\infty}
{{{\rm e}^{-y^2}}\over{(u-y)^2+a^2}}dy.
\label{eq:24}
\end{equation}
With the help of this function, the absorption coefficient per atom, $\alpha_\lambda$, is expressed as
\begin{equation}
\alpha_\lambda =
{{e^2}\over{4\pi\varepsilon_0 m_{\rm e} c^2}} \pi^{1/2} {{\lambda_0^2 f}\over{\lambda_{\rm D}}} H(u,a),
\label{eq:25}
\end{equation}
where $\lambda_{\rm D}$ denotes the Doppler width corresponding to the variance of the Maxwellian velocity distribution of thermal motion, 
$u=(\lambda-\lambda_0)/\lambda_{\rm D}$, $a=\Gamma \lambda_0^2/(4\pi c\lambda_{\rm D})$,  ${e}$ and $m_{\rm e}$ are electronic charge and mass, respectively, and $f$ denotes the oscillator strength, which is different for each atomic level. Most $f$ values are determined from empirical laboratory measurements. Only in less complicated cases, such as hydrogen lines, can theoretical calculations be done.

\subsubsection{Damping parameter}
\label{sec:3.2.4}
\noindent
The damping parameter, $\Gamma$, is expressed as the sum of that of radiation damping, $\gamma$, the quadratic Stark effect, $\gamma_4$, and the van der Waals effect, $\gamma_6$.
As for radiation damping, we adopt the classical dipole emission theory and set
\begin{equation}
\gamma={{8\pi}\over{3}}{{e^2}\over{4\pi \varepsilon_0 m_{\rm e}c\lambda_0^2}},
\label{eq:26}
\end{equation}
where $\lambda_0$ is the center wavelength of the line. 
We evaluate $\gamma_4$ and $\gamma_6$, following \citet{Gray05}, by
\begin{equation}
\log\gamma_4 = 18+{{2}\over{3}}\log C_4 + \log P_{\rm e} -{{5}\over{6}}\log T 
\label{eq:27}
\end{equation}
and
\begin{equation}
\log\gamma_6 = 19+0.4\log C_6 + \log P_{\rm gas} -0.7\log T, 
\label{eq:28}
\end{equation}
respectively, where $P_{\rm e}$ and $P_{\rm gas}$ are in units of Pa.  
Here,
$C_4$ is a constant, but dependent on the line, and $C_6$ is given by
\begin{equation}
C_6 = 0.3\times{10}^{-30}\left[(I-\chi-\chi_\lambda)^{-2} - (I-\chi)^{-2}\right],
\label{eq:29}
\end{equation}
where $I$ denotes the ionization energy, $\chi$ is the excitation potential, and $\chi_\lambda=hc/\lambda$. 
$I$, $\chi$, and $\chi_\lambda$ are in units of eV.

\subsubsection{Linear Stark effect for hydrogen lines}
\label{sec:3.2.5}
Hydrogen lines are different from other elements' ones in the sense that the atomic structure of hydrogen is subject to the linear Stark effect. The complete hydrogen Stark broadening is the weighted sum of the individual Stark components with each showing the Holtsmark distribution. We take into account of the linear Stark effect in the case of hydrogen lines following \citet{Gray05}. The $f$-values of the unshifted Stark components are denoted as $f_0$ and the $f$-values of the shifted components are denoted as $f_{\pm}$. These values are given in table\,\ref{tab:3}.
\begin{table}
\caption{Hydrogen oscillator strengths. Adopted from \citet{Gray05}.}
\label{tab:3}
\begin{center}
\begin{tabular}{clll}
\hline
Line & $f_0$ & $f_{\pm}$ & $f$  \\ \hline
H\,$\alpha$ & 0.248689 & 0.392058 & 0.640742 \\
H\,$\beta$ & 0.00 & 0.119321 & 0.119321 \\
H\,$\gamma$ & 0.007014 & 0.037656 & 0.044670 \\
H\,$\delta$ & 0.00 & 0.022093 & 0.022093 \\ \hline
\end{tabular}
\end{center}
\end{table}

The approach of \citet{Stehle94} is used for H\,$\beta$, while that of
\citet{Vidal_etal73} is used for the other Balmer lines.
With the help of these Stark profiles, we calculate the absorption coefficient for hydrogen by 
\begin{eqnarray}
\alpha_\lambda &=&
{{e^2}\over{4\pi\varepsilon_0 m_{\rm e} c^2}} \pi^{1/2}{{\lambda_0^2}\over{\lambda_{\rm D}}} 
\nonumber \\
&& 
\times
\left[ f_{\pm} {{S(\Delta\lambda/E_0)}\over{E_0}}*H(u,a) + f_0H(u,a)\right],
\label{eq:30}
\end{eqnarray} 
where * means convolution.
Here, 
\begin{equation}
E_0=\left({{4\pi n}\over{3}}\right)^{2/3} {{e}\over{4\pi\varepsilon_0}},
\label{eq:31}
\end{equation}
where $n$ denotes the number density of ions.  

\subsection{Scattering coefficient}
\label{sec:3.3}
\noindent
The contribution of Rayleigh scattering is substantially large at shorter wavelengths ,such as $\lambda \ll 300\,{\rm nm}$. But, for the optical region, it is negligibly small. We therefore only have to consider Thomson scattering by free electrons. The scattering coefficient is then independent of wavelength, and is given by
\begin{equation}
\sigma_\lambda = {{8\pi}\over{3}}
\left( {{{e}^2}\over{4\pi\varepsilon_0 m_{\rm e}c^2}} \right)^2
{{n_{\rm e}}\over{\rho}},
\label{eq:32}
\end{equation}
where 
$n_{\rm e}$ denotes the electron number density.

\section{Check of our numerical code}
\label{sec:4}
\subsection{Intrinsic line profiles}
\label{sec:4.1}
\noindent
In order to check our numerical code, we calculated first the line profile of the Balmer lines by using a solar model \citep{JCDetal96}. 
The upper panel of figure\,\ref{fig:2} shows the computed line profile of H\,$\alpha$ line. 
In this procedure, the effect of rotational broadening was not taken into account, and micro-turbulence was ignored. The values of $\gamma_4$ and $\gamma_6$ were adjusted so that the wing part of the computed profile fit well to the observed one (e.g., \citet{Babcock_Moore47}). 
We confirmed how well the computational programs developed here do work, though there remains some discrepancy. Fitting is substantially improved in the case of H\,$\beta$ and H\,$\gamma$. 
\begin{figure}[htb]
\begin{center}
\includegraphics[width=0.95\linewidth,angle=0]{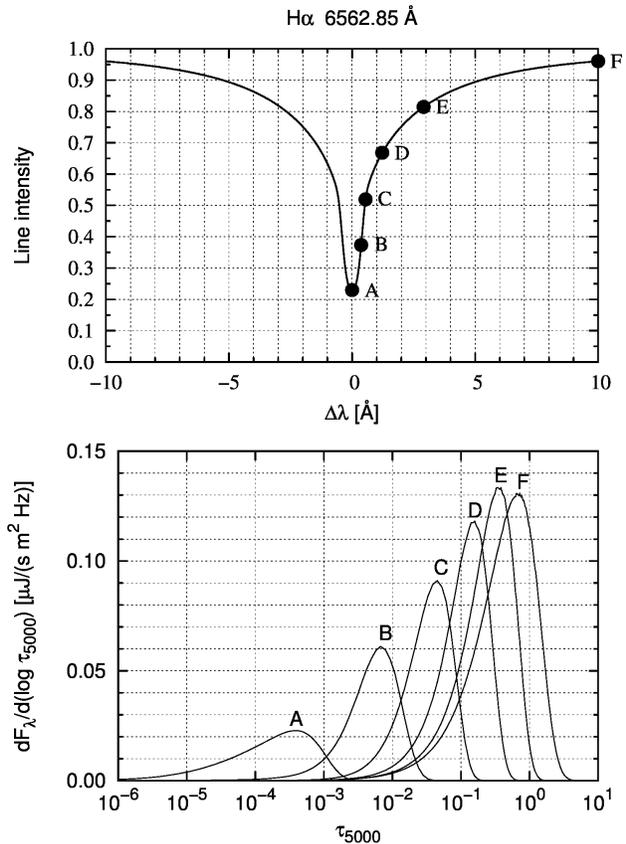}
\end{center}
\caption{Upper: Computed line profile of the H\,$\alpha$ line in a wide range. The ordinate is the flux normalized with the continuum flux, and the abscissa is the wavelength measured from the line center, in units of {\AA}.
Lower: Contribution functions for H\,$\alpha$ line for the solar model. These are the integrand of the integral with respect to $\tau_{\lambda z}$ in equation (\ref{eq:9}), evaluated at the wavelengths, which are marked along the computed line profile by dots in the upper panel.  
The abscissa is the optical depth for $\lambda=5000$\,\AA, $\tau_{5000}$. 
}
\label{fig:2}
\end{figure}

How much does each layer of the atmosphere contribute to formation of the spectral line?
This can be seen by the integrand of the integral with respect to $\tau_{\lambda z}$, near the disk center, in equation (\ref{eq:9}). It is often called the contribution function. The lower panel of figure\,\ref{fig:2} shows the contribution functions for H\,$\alpha$ line for the solar model. These are the integrand of the integral with respect to $\tau_{\lambda z}$ in equation (\ref{eq:9}), evaluated at the wavelengths, which are marked along the computed line profile by dots in the upper panel of figure\,\ref{fig:2}.  
The abscissa is the optical depth for $\lambda=5000$\,{\AA}, $\tau_{5000}$. The contribution function for the line center concentrates around $\tau_{5000} \simeq 10^{{-3}\sim{-4}}$, which corresponds to $300\sim 500$\,km above the photosphere.

\subsection{The LPV caused by radial pulsation}
\label{sec:4.2}
\noindent
Let us see the influence of pulsation on the line profile of H\,$\alpha$ line thus calculated. The current purpose is to see how much difference in the line profile variation is caused by the difference in the vertical wavelength of motion. For this reason, we assume radial pulsation of a non-rotating star, in which case no lateral motion exists. 

We consider three cases: (i) a standing wave with a  constant pulsation amplitude, which is regarded as representing a case of very long vertical wavelength; (ii) a standing wave with the wavelength of the order of the thickness of the line forming layer; (iii) an upwardly running wave with the wavelength of the order of the thickness of the line forming layer.   
\begin{figure*}[htb]
\begin{center}
\vspace{0.2cm}
\includegraphics[width=0.9\linewidth,angle=270]{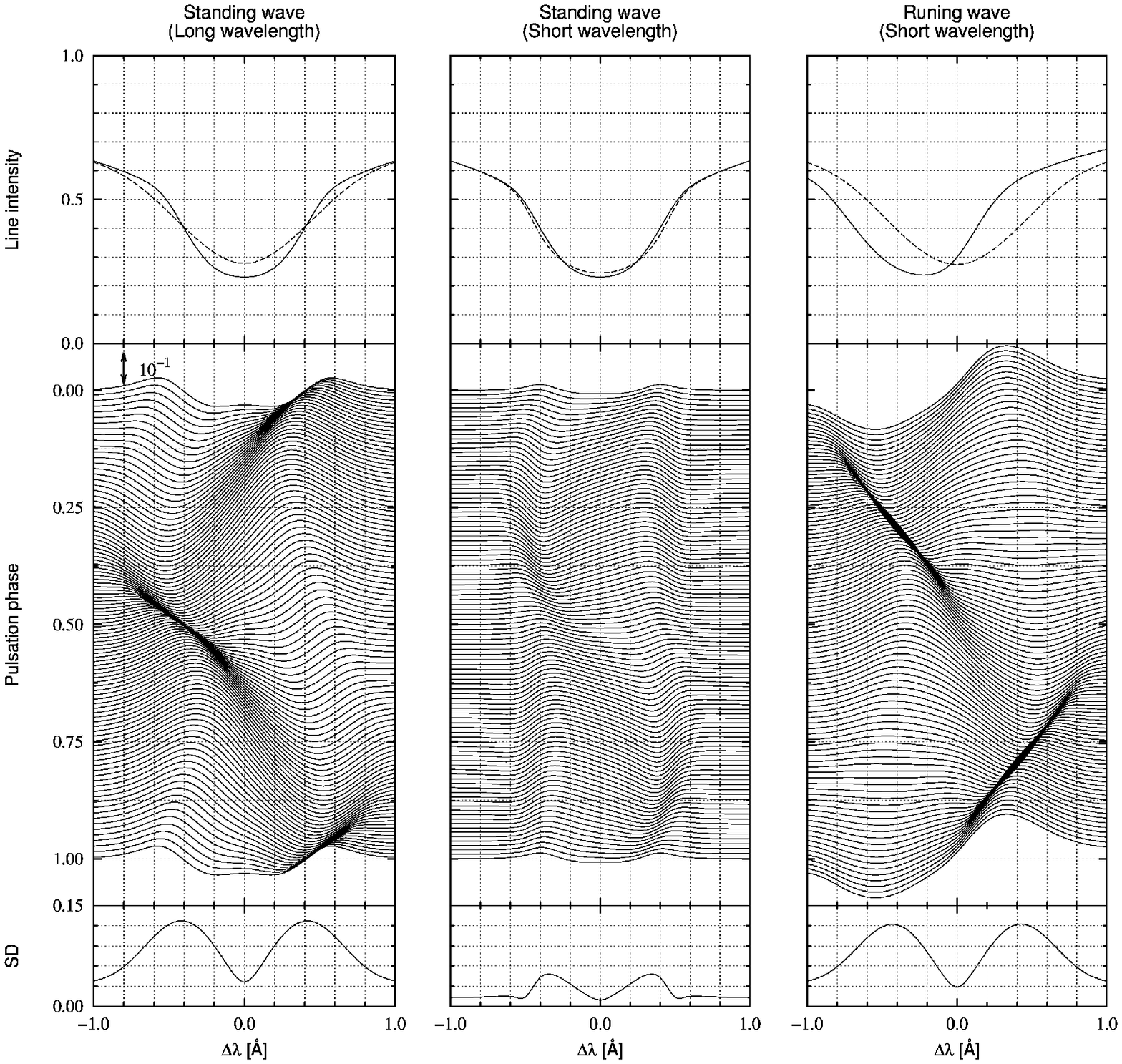}
\end{center}
\caption{Theoretically expected line-profile variation (LPV) of H\,$\alpha$ line produced by a sinusoidally oscillating radial mode of a nonrotating star. 
Left: Pulsation velocity amplitude was assumed to be spatially constant. This is regarded as representing a case of very long vertical wavelength. 
Middle: Oscillation was assumed to be a standing wave with the wavelength of the order of the thickness of the line forming layer. 
Right: Oscillation was assumed to be an upwardly running wave with the wavelength of the order on the thickness of the line forming layer.  
In each panel, the differences between the individual profiles at a sequence of 100 phases and their mean are stacked with phase increasing downwards. 
A short bar in the left panel indicates the length corresponding to $10\,\%$ of the continuum intensity.
Above them are their mean (dashed line) and the line profile at the phase zero (solid line), and below them is plotted their standard deviation. The abscissa is the wavelength from the line center of the intrinsic profile, in units of {\AA}. The left-hand ordinate scales refer to the phase of the oscillation. 
}
\label{fig:3}
\end{figure*}

The line profile variation is most clearly seen in the time series of the residuals that remain after subtracting from each line profile its temporal mean. In the left panel of figure\,\ref{fig:3} is displayed the line-profile variation (LPV) of the case (i), obtained from the current calculation using a solar model. Plotted is the deviation of the integrated radiant intensity from its mean as a function of wavelength for a sequence of values of the pulsation phase extending over one complete period. Also, the line profile at the phase zero is displayed to compare with the mean profile averaged over one period. Since the pulsation amplitude is assumed to be spatially constant in this case, the line-forming layer moves up and down as a whole. Consequently, the spectral line shifts from blue to red and back again, while keeping its symmetric profile. This behavior can be clearly seen in the motion of the bisector of the line profile, which is displayed in the left panel of figure\,\ref{fig:4}. The bisector is almost always a straight line, and it simply moves from side to side with time 
\footnote{The small deviation from a straight line reflects the spherical effect of stratification. The angle between the normal to an atmospheric layer and the line of sight gradually becomes slightly larger with height of the layer. Hence, the Doppler shift is slightly larger with line intensity even if the pulsation velocity amplitude is a constant.}. 
\begin{figure*}[ht]
\begin{center}
\vspace{0.2cm}
\includegraphics[width=0.51\linewidth,angle=0]{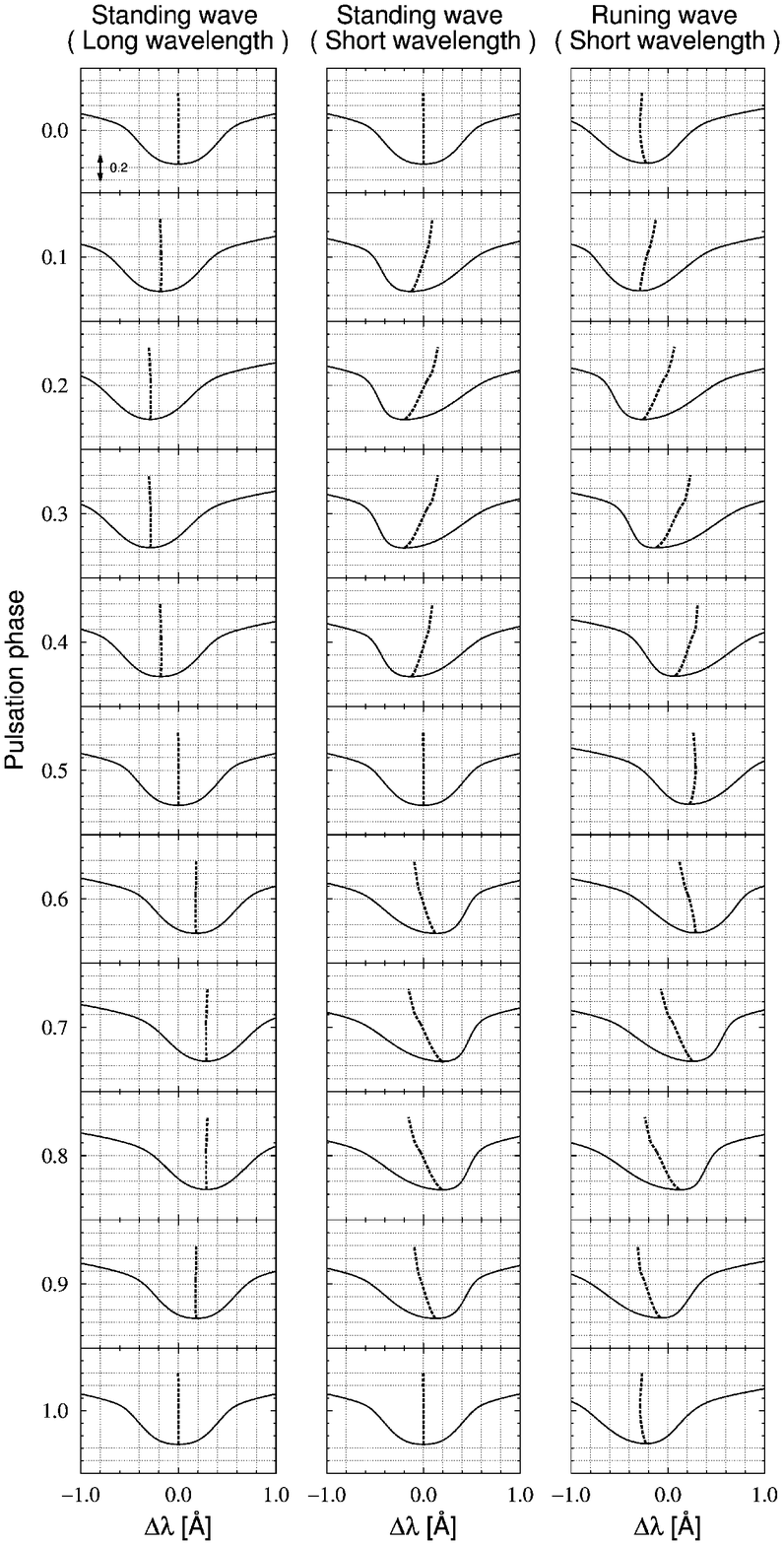}
\end{center}
\caption{
Theoretically expected variation in the bisector of line profiles during a one complete period. 
In each panel, the line profiles (solid lines) and the bisectors (dashed lines) at the pulsation phases 0.0, 0.1, ..., 1.0 are stacked with phase increasing downwards. The abscissa is the deviation from the line center of the intrinsic profile, and the ordinate of each diagram at a fixed pulsation phase is the line intensity normalized with the continuum level. 
Left: Case of a constant velocity amplitude. 
Middle: Case of a standing wave with the wavelength of the order of the thickness of the line forming layer. 
Right: Case of an upwardly running wave with the wavelength of the order of the thickness of the line forming layer. 
A short bar in the left top panel indicates the length corresponding to $20\,\%$ of the continuum intensity. 
}
\label{fig:4}
\end{figure*}

The middle panel of figure\,\ref{fig:3} displays the LPV seen in the case of (ii). The difference from the case (i) would be hardly perceptible in this form, but it is clearly seen in the motion of the bisector, which is shown in the middle panel of figure\,\ref{fig:4}. The vertical wavelength of pulsation has been assumed in this case to be as short as the thickness of the line-forming layer. As a result, the velocity varies with height, and the line shape no longer has symmetry. Furthermore, there is a nodal surface at a certain level. These features are well recognized in the middle panel of figure\,\ref{fig:4}.  
The right panel of figure\,\ref{fig:3} displays the LPV seen in the case of a  running wave, propagating upward in the atmosphere. The running wave feature is clearly seen in the motion of bisector (the right panel of figure\,\ref{fig:4}). On the contrary to the case of a standing wave, the line depth at which the bisector crosses the line center of the intrinsic line profile does not stay still, but shifts downward with time.

\begin{figure}[tb]
\begin{center}
\includegraphics[width=0.95\linewidth,angle=0]{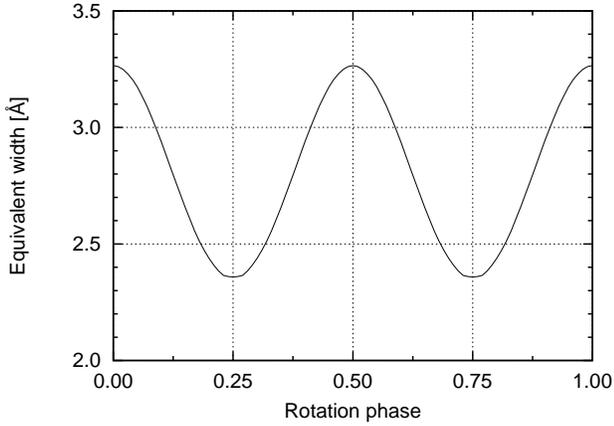}
\end{center}
\caption{Variation in the equivalent width of H\,$\alpha$ line in the case that lateral distribution of hydrogen in the atmosphere is proportional to $|\cos\theta |$, where $\theta$ denotes the colatitude with respect to the magnetic axis, which is assumed to be perpendicular to the rotation axis assumed to be perpendicular as well to the line of sight ($i=90^\circ$ and $\beta=90^\circ$). 
The rotation phase zero is the phase when the magnetic axis is oriented toward the line of sight.
The ordinate is in units of {\AA}.}
\label{fig:5}
\end{figure}

\subsection{Influence of lateral chemical inhomogeneity and rotation of the star}
\label{sec:4.3}
\noindent
If the chemical distribution in the stellar atmosphere is laterally homogeneous, the effect of stellar rotation on a stellar spectrum is to broaden spectral lines. The amount of broadening depends on the rotation velocity and the the inclination of the rotation axis to the line of sight, while the equivalent width is kept independent of rotation. 
As explained in section\,\ref{sec:1}, the atmospheres of magnetic Ap stars are laterally inhomogeneous and some chemical elements are selectively levitated near the magnetic polar region. To see the effect of such lateral inhomogeneity upon LPV, we assume here that hydrogen is slightly concentrated in the magnetic polar region. 
The number density of hydrogen is supposed to vary proportionally to $|\cos\theta|$, where $\theta$ is the colatitude with respect to the magnetic axis. Naturally enough, the equivalent width of H\,$\alpha$ line is dependent on the aspect angle. As a consequence, if the magnetic axis is misaligned with the rotation axis of the star, the equivalent width of hydrogen lines synchronously varies with the rotation. To see the largest effect, we demonstrate here, in figure \ref{fig:5}, the most extreme case, in which the line opacity is proportional to $|\cos\theta|$, though it is implausible in the case of H\,$\alpha$ line. 
\begin{figure}[thb]
\begin{center}
\includegraphics[width=0.50\linewidth,angle=0]{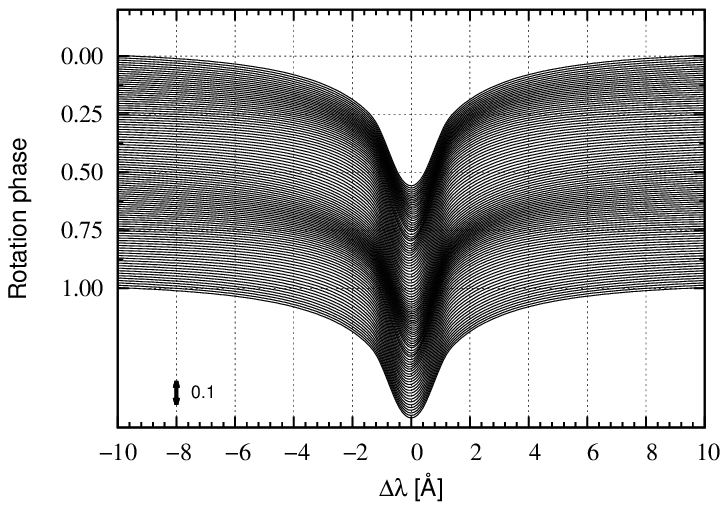}
\includegraphics[width=0.50\linewidth,angle=0]{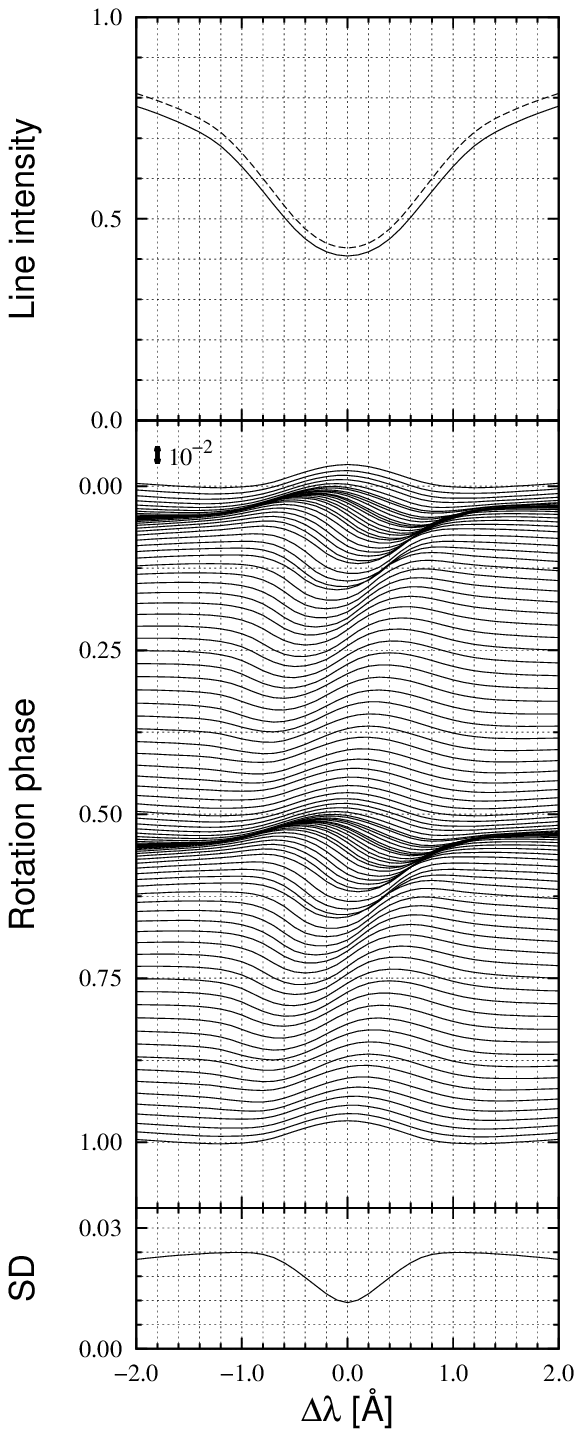}
\end{center}
\caption{Theoretically expected LPV of H\,$\alpha$ line in the case that hydrogen is slightly concentrated near the magnetic polar region in the atmosphere.
Upper: The individual profiles at a sequence of 100 phases with phase increasing downwards.
Lower: The differences between the individual profiles at a sequence of 100 phases, and their mean are stacked with phase increasing downwards; above them are their mean (dashed line) and the line profile at the phase zero (solid line), and below them is plotted their standard deviation. 
In each panel, the abscissa is in units of {\AA}, and the left-hand ordinate scales refer to the rotation phase. 
A short bar in each panel indicates the length corresponding to $1\,\%$ of the continuum intensity. 
}
\label{fig:6}
\end{figure}
  
Figure\,\ref{fig:6} shows the line profiles in the case of $i=90^\circ$ and $\beta=90^\circ$ at a sequence of 100 rotation phases (the upper panel) and the differences between the individual profiles and their mean (the lower panel).
When one magnetic polar region is coming to appear on the visible hemisphere while the other polar region is coming to disappear from there, the apparent distribution of the element over the visible hemisphere becomes conspicuously inhomogeneous. As a consequence, the line shape becomes asymmetric at such a rotational phase. 

\section{Calculation of line-profile variation in A-type stars}
\label{sec:5}
\subsection{Equilibrium model for A-type stars}
\label{sec:5.1}
\noindent
Based on the numerical technique explained in the previous sections, let us proceed to the case of Ap stars. We choose a 2\,$M_\odot$ star at the zero age as an equilibrium model. The star is assumed to be chemically homogeneous throughout the star, and the hydrogen mass ratio, $X$, and the metal mass ratio, $Z$, are assumed to be $0.73$ and $0.02$, respectively, and the chemical composition of the solar abundance is assumed. The equilibrium model of the star is constructed with a program originally written by \citet{Paczynski70}. The opacity table in the program is, however, updated, and we adopt the OPAL \citep{Iglesias_etal96} for a wide range of temperature and supplement the low temperature range ($\log T \leq 4.0$) with \citet{Ferguson_etal05}. 
The global physical parameters of the model are summarized in table\,\ref{tab:4}. 
\begin{table}[htdp]
\caption{Physical properties of the equilibrium model.}
\begin{center}
\begin{tabular}{cccccc}
\hline
$M/M_\odot$ & $R/R_\odot$ & $L/L_\odot$ & $T_{\rm eff}$ & $X$ & $Z$  \\ \hline
$2.0$ & $1.64$ & $14.7$ & $8800$\,K & $0.73$ & $0.02$ \\ \hline
\end{tabular}
\end{center}
\label{tab:4}
\end{table}

Although the atmospheric structure of Ap stars has not been definitely established, it may be substantially deviate from the standard one due to the strong magnetic fields and peculiar chemical compositions. Following \citet{Shibahashi_Saio85}, we have adopted an analytic form of the $T$-$\tau$ relation for the atmosphere, which was fitted to Kurucz's (1979) model atmosphere:
\begin{eqnarray}
T^4&=&{{3}\over{4}}T_{\rm eff}^4 
[\tau+0.9052-0.3367\exp(-2.54\tau)
\nonumber \\
& & -0.1645\exp(-300\tau)-0.1440\exp(-10000\tau).
\label{eq:33}
\end{eqnarray}
The top panel of figure\,\ref{fig:7} displays the relation between the optical depth at $\lambda=5000$\,\AA\, and the geometrical depth from the top layer for the present model. 

The sound speed profile in the atmosphere is displayed in the bottom panel of figure\,\ref{fig:7}. The sound speed is about $10\,{\rm km}\,{\rm s}^{-1}$ at the photosphere, and becomes as slow as $7\,{\rm km}\,{\rm s}^{-1}$ in the high atmosphere. In the case of acoustic modes with a period of $\sim 600\,{\rm s}$, the wavelength is as short as the thickness of the atmosphere. 

\subsection{Ionization degrees of hydrogen and neodymium ions}
\label{sec:5.2}
\noindent
It should be noted here that, contrary to the solar case, the density in the high atmosphere is so low that hydrogen is ionized there ($\tau_{5000} > 10^{-5}$), as shown in figure\,\ref{fig:8}. Consequently, the H\,$\alpha$ line in A-type stars is more sensitive to the very outer layer than in the case of the Sun. This is clearly seen in the contribution function shown in figure\,\ref{fig:9}, which should be compared with figure\,\ref{fig:2},  displaying the contribution function for the solar model. 
\begin{figure}[htb]
\begin{center}
\includegraphics[width=0.6\linewidth,angle=0]{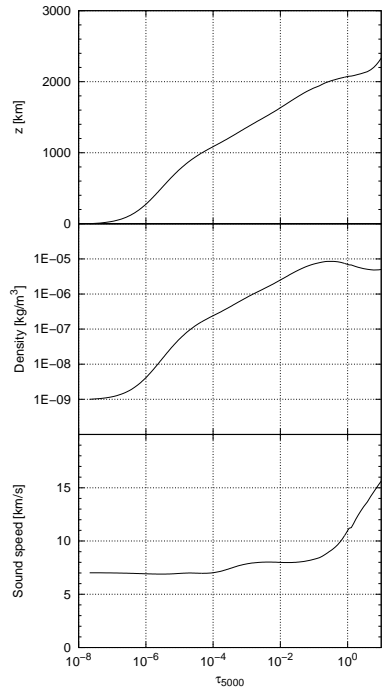}
\end{center}
\caption{Top: Depth from the top layer of the present model as a function of the optical depth at $\lambda=5000$\,{\AA}. 
Middle: Density profile in the atmosphere as a function of the optical depth at $\lambda=5000$\,\AA.
Bottom: Sound speed in the atmosphere as a function of the optical depth at $\lambda=5000$\,{\AA}.
}
\label{fig:7}
\end{figure}

\begin{figure}[tbh]
\begin{center}
\vspace{0.8cm}
\includegraphics[width=0.81\linewidth,angle=0]{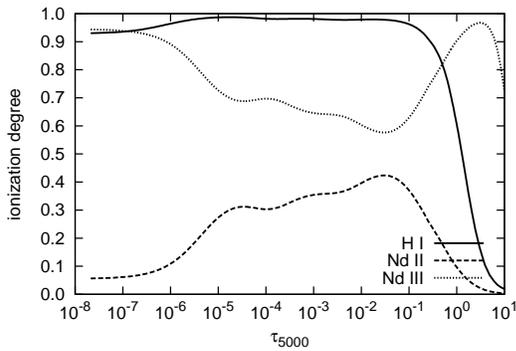}
\end{center}
\caption{Ionization degrees of neodymium and of hydrogen in the present model of an A-star.  Plotted as a function of the optical depth $\tau_{5000}$. 
}
\label{fig:8}
\end{figure}

\begin{figure}[hbt]
\begin{center}
\includegraphics[width=0.81\linewidth,angle=0]{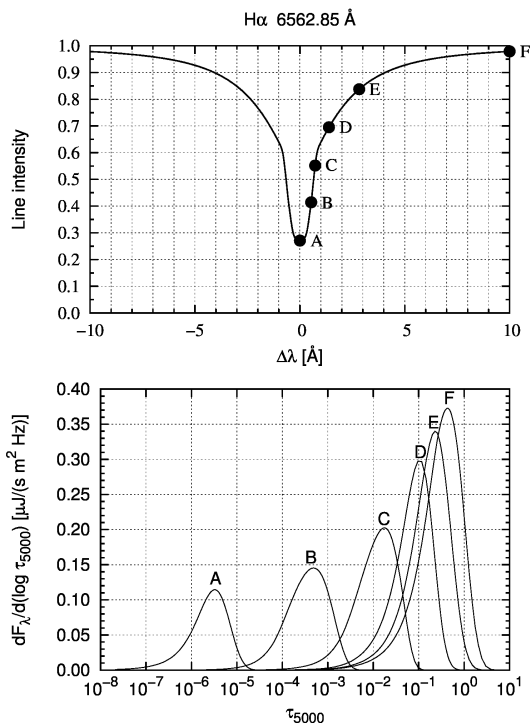}
\end{center}
\caption{
Upper: Computed line profile of the H\,$\alpha$ line in a wide range. The abscissa is the wavelength measured from the line center, in units of {\AA}.
Lower: Contribution functions for H\,$\alpha$ line for present model of an A-star. These are the integrand of the integral with respect to $\tau_{\lambda z}$ in equation (\ref{eq:9}), evaluated at the wavelengths, which are marked along the computed line profile by dots in the upper panel.  
The abscissa is the optical depth for $\lambda=5000$\,\AA, $\tau_{5000}$. 
}
\label{fig:9}
\end{figure}

Recent high-speed spectroscopic observations of oscillations of roAp stars have revealed that spectacular line-profile variation are seen in the spectrum lines of Pr\emissiontype{III} and Nd\emissiontype{III}. It is these observations that have motivated the work reported in this paper. We then pay attention to Nd\emissiontype{III}. Figure\,\ref{fig:8} displays the ionization degrees of neodymium ions, as well as that of hydrogen, in the present A-star model. The ionization degree of Nd\emissiontype{III} goes up with height after reaching a local minimum at a few hundred kilometers above the photosphere, and almost all of neodymium is Nd\emissiontype{III} in the high atmosphere. This is the reason why Nd\emissiontype{III} lines show much spectacular line-profile variation than Nd\emissiontype{II} lines. In these calculations, the atomic data of Nd were taken from a database offered by US National Institute of Standards and Technology (http://physics.nist.gov). However, the statistical weight of Nd\emissiontype{IV} was not available there. We took the value from Kurucz's ATLAS and set $\log {\sl g}=1.8$ for any energy levels. 

\subsection{Contribution functions}
\label{sec:5.3}
\noindent
The abundance of neodymium involved in Ap stars seems to generally be much larger than the solar case, in which $\log (N_{\rm Nd}/N_{\rm H}) +12 = 1.42\pm 0.04$ \citep{Asplund2009}, and some stars show overabundance of 4 dex \citep{Adelman73}. In many cases, the observed Nd\emissiontype{III} lines with low excitation energies are strong, though Nd\emissiontype{II} lines are much weaker. This implies that neodymium is mainly concentrated in a thin layer above $\tau_{5000} < 10^{-5}$. \citet{Ryabchikova_etal02} modeled the spectra of one of Ap stars, $\gamma$\,Equ, by supposing $\log(N_{\rm Nd}/N_{\rm H})\simeq -4$ for $\log\tau < -9$ under the assumption of LTE, where $N_{\rm Nd}/N_{\rm total}$ denotes the fraction of neodymium. \citet{Mashonkina_etal05} modeled the same star with the assumption of NLTE by supposing $\log(N_{\rm Nd}/N_{\rm H})\simeq-7$ for $\log\tau < -4$.

\begin{figure}[bht]
\begin{center}
\includegraphics[width=0.82\linewidth,angle=0]{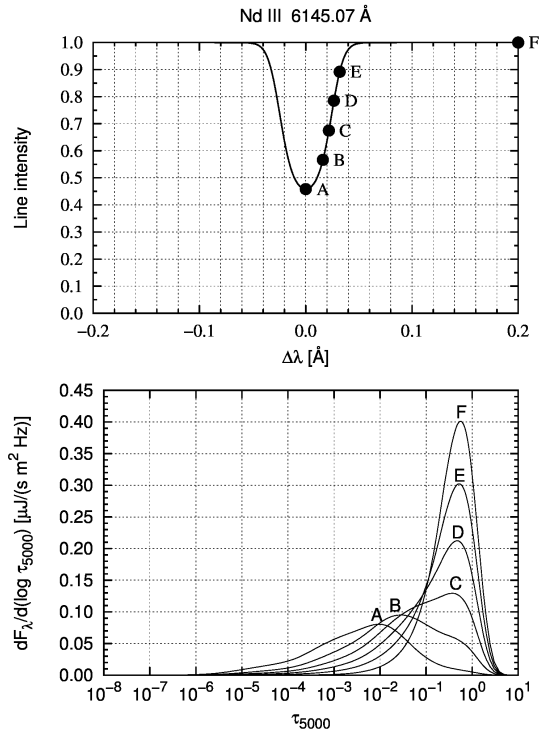}
\end{center}
\caption{
Upper: Line profile of Nd\emissiontype{III} $\lambda\, 6145.07$\,{\AA} line for the current model with $\log (N_{\rm Nd}/N_{\rm H}) = -8$. 
The abscissa is the wavelength measured from the line center, in units of {\AA}.
Lower: Contribution functions for the above line. These are the integrand of the integral with respect to $\tau_{\lambda z}$ in equation (\ref{eq:9}), evaluated at the wavelengths marked along the computed line profile by dots in the upper panel.  
The abscissa is the optical depth for $\lambda=5000$\,\AA, $\tau_{5000}$. 
}
\label{fig:10}
\end{figure}

In this paper, we assume, however, the vertically uniform distribution of elements for the sake of simplicity, for our current main purpose is to demonstrate the usefulness of analyses of LPV as an information source of oscillation and atmospheric structures of roAp stars, rather than to model the individual star in detail.   
The contribution functions for Nd\emissiontype{III} $\lambda\, 6145.07$\,{\AA} line in the case of $\log(N_{\rm Nd}/N_{\rm H})= -8$ and in the case of $\log(N_{\rm Nd}/N_{\rm H})= -5$ are shown in figures\,\ref{fig:10} and \ref{fig:11}, respectively. The oscillator strength was taken from \citet{Cowley_Bord98}. The effect of rotation of the star was ignored in these calculations.  
\begin{figure}[tbh]
\begin{center}
\includegraphics[width=0.82\linewidth,angle=0]{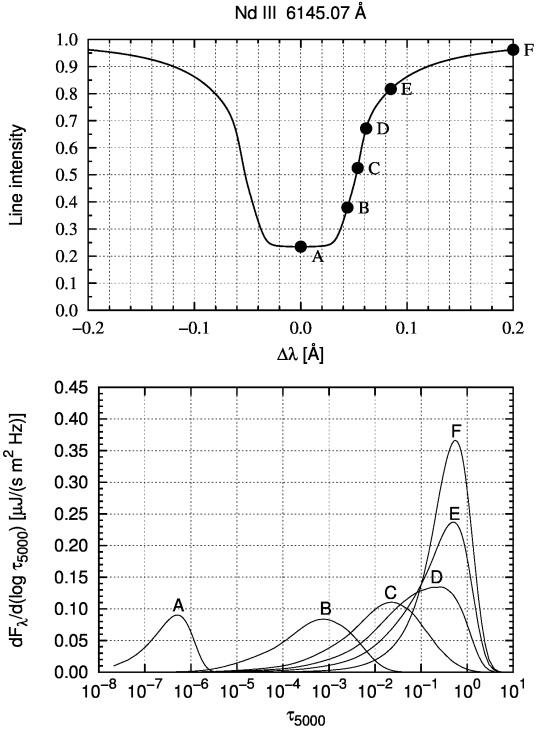}
\end{center}
\caption{
Same as figure\,\ref{fig:10}, but for the case of $\log (N_{\rm Nd}/N_{\rm H}) = -5$.
}
\label{fig:11}
\end{figure}

\subsection{Line-profile variation due to nonradial oscillations in A-type stars}
\label{sec:5.4}
\noindent
As explained in section\,\ref{sec:1}, the general features of pulsations in roAp stars are essentially well explained in terms of high order, axisymmetric dipole p-modes, of which the symmetry axis is aligned to the magnetic axis of the star, which is inclined to the rotation axis of the star. In this subsection, we simulate line-profile variation of the H\,$\alpha$ line and Nd\emissiontype{III} $\lambda\,6145.07$\,{\AA} due to such oscillation modes in roAp stars by using the equilibrium model described in the previous subsection. Both of the elements are assumed to slightly concentrate in the magnetic polar region, distributed varying proportionally to $\cos^2\theta$, where $\theta$ denotes the polar angle measured from the magnetic axis. The following simulations are for the case of line opacity being proportional to $\cos^2\theta + 1/2$. We treat the inclination angle between the rotation axis of the star and the symmetry axis of pulsation as a free parameter, $\beta$. Also the angle between the line-of-sight and the rotation axis of the star is treated as another free parameter, $i$. 
The aspect angle of the symmetry axis of pulsation, $\alpha$, gradually changes with time due to rotation of the star, as follows:
\begin{equation}
\cos\alpha=\cos  i\cos\beta +\sin i\sin\beta\cos\Omega t,
\label{eq:34}
\end{equation}
where $\Omega$ denotes the rotational frequency. 
We will show the theoretically calculated line profiles at every 0.125 rotation phase. 

\vspace{0.3cm}
\subsubsection{Standing eigenmode}
\label{sec:5.4.1}
\noindent
We consider here a standing eigenoscillation of the A-star model. For the sake of simplicity, the oscillation is assumed to be adiabatic.  
Also, we ignore the effects of rotation of the star and treat the star as being spherically symmetric, and deal with only low-degree, high-order p-modes, of which period is 600\,s; ---in a typical period range of roAp stars. In this case, the Eulerian perturbation to the gravitational potential is negligibly small, and then the basic equations governing the linear, adaiabatic oscillations of a star are well described with only equations of conservations of momentum, mass, and entropy; they are reduced to a form expressed with only the velocity perturbation, $\mbox{\boldmath$v$}$;
\begin{eqnarray}
\rho{{\partial^2 \mbox{\boldmath$v$}}
\over{\partial t^2}}
=
&&\nabla(\mbox{\boldmath$v$}\cdot \nabla p)
+
\nabla(\rho c^2\nabla\cdot\mbox{\boldmath$v$})
\nonumber\\
&-&
{{1}\over{\rho}}(\mbox{\boldmath$v$}\cdot\nabla\rho)\nabla p
-(\nabla\cdot\mbox{\boldmath$v$})\nabla p,
\label{eq:35}
\end{eqnarray}
where $p$ and $\rho$ denote the pressure and the density, respectively, and $c$ means the sound speed. We solve numerically equation (\ref{eq:35}) with a set of proper boundary conditions to compute the eigenfunctions of linear, adiabatic, dipole oscillations of the A-star model, following the procedure described in \citet{Unno_etal89}.
The amplitude is arbitrarily chosen in the linear calculation. We choose it so that the maximum velocity is still subsonic; $|\mbox{\boldmath$v$}| = 7\,{\rm km}\,{\rm s}^{-1}$ at the top layer of the atmosphere.
  
\begin{figure*}[tbh]
\begin{center}
\vspace{0.4cm}
\includegraphics[width=0.5\linewidth,angle=90]{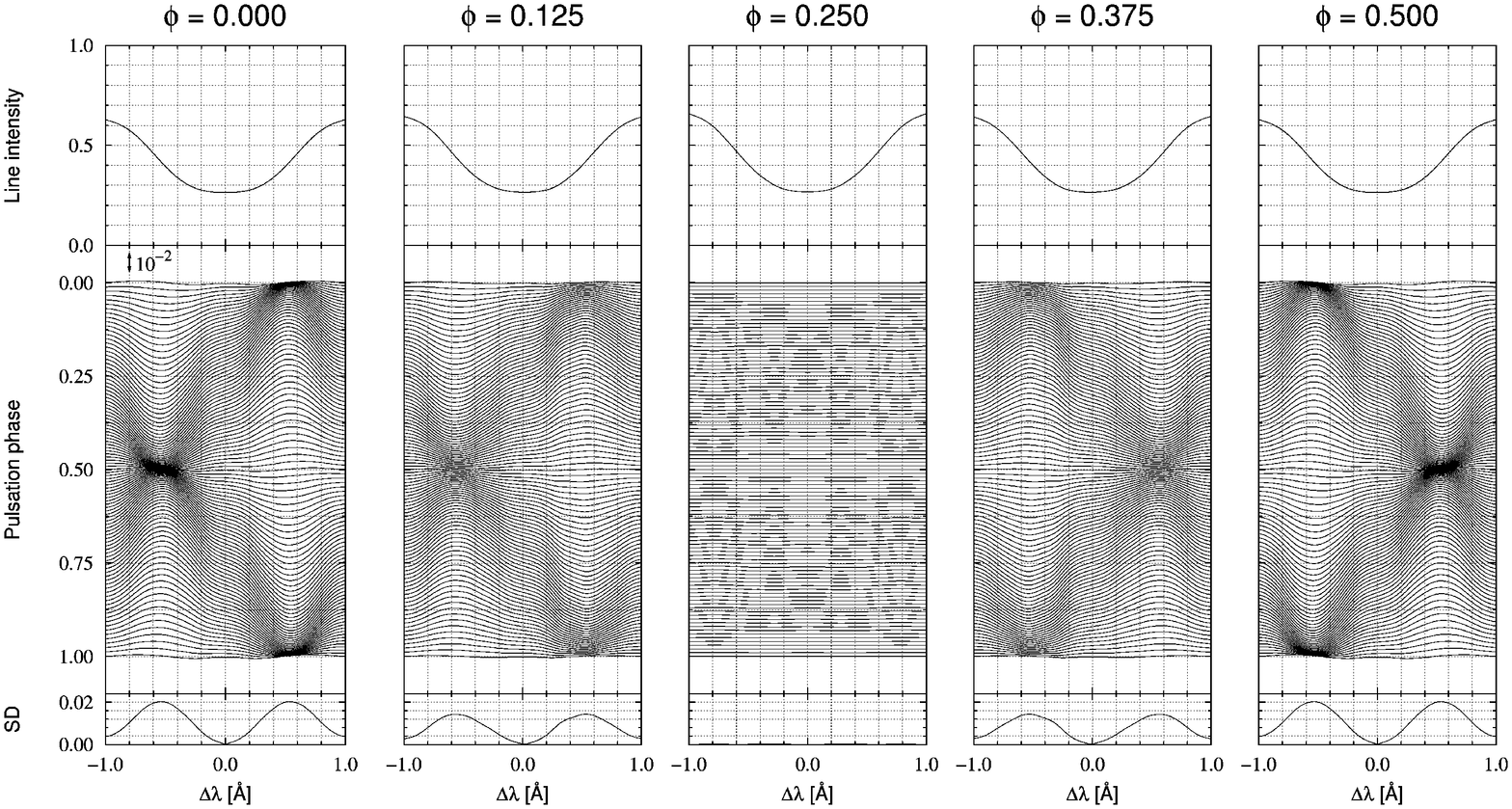}
\end{center}
\caption{
Theoretically expected LPV of the H\,$\alpha$ line produced by an axisymmetric dipole $(l=1, m=0)$ mode at a sequence of  the rotation phase $\phi=$0.0, 0.125, 0.25, 0.375, and $0.5$ (from left to right). The rotation axis is assumed to be inclined with respect to the line-of-sight by $90^\circ$, and the symmetry axis of pulsation is assumed to be inclined with respect to the rotation axis of the star by $90^\circ$. 
In each panel, the differences between the individual profiles at a sequence of 100 pulsation phases and their mean are stacked with phase increasing downwards; above them is their mean, and below them is plotted their standard deviation. The abscissa is the wavelength measured from the line center in units of {\AA}. The left-hand ordinate scales refer to the phase of the oscillation, in units of the pulsation period.  
A short bar in the left panel indicates the length corresponding to $1\,\%$ of the continuum intensity. 
}
\label{fig:12}
\end{figure*}

In the outer envelope of a star, the effect of sphericity becomes unimportant. To understand the oscillation feature there,  it is instructive to consider a simple solution of plane waves of the form 
\begin{equation}
\mbox{\boldmath$v$} \propto \exp\left( -{{z}\over{2H_\rho}} \right)
\exp(i\mbox{\boldmath$k$}\cdot\mbox{\boldmath$x$}+i\omega t)
\label{eq:36}
\end{equation}
with an assumption of $k_z \gg H_\rho$, where $H_{\rho}\equiv -dz/d\ln\rho$ is the density scale height and $\omega$ is the frequency.
The exponentially growing factor with height in equation (\ref{eq:36}) arises so as to conserve wave energy in vertical direction since the density in atmosphere decreases with height.
Substituting equation (\ref{eq:36}) into equation (\ref{eq:35}), we obtain a local dispersion relation, which gives the vertical wavenumber $k_z$ for a given set of the horizontal wavenumber and the frequency. For high-order p-modes, the dispersion relation is reduced to
\begin{equation}
k_z^2 (z)= c^{-2}(z)\, \left\{\omega^2- \omega_{\rm c}^2(z)\right\},
\label{eq:37}
\end{equation}
where
\begin{equation}
\omega_{\rm c}^2 (z)\equiv {{c^2}\over{4H_\rho^2}}\left(1+2{{dH_\rho}\over{dz}}\right)
\label{eq:38}
\end{equation}
denotes the squared critical frequency.
Equation (\ref{eq:37}) means that the oscillation is evanescent and the wave energy, $\rho |\mbox{\boldmath$v$}|^2$, decreases monotonically with height if $\omega^2 < \omega_{\rm c}^2(z)$. The latter condition is required for the wave to be standing wave. 
Since what we are now considering are standing eigenmodes, they are trapped waves that are reflected at the two boundaries of the system. Hence, they become evanescent above the photosphere, and then have no node in the line-forming layer. However, the frequencies are so close to the critical frequency near the surface that the $e$-folding scale of attenuation of evanescent wave energy is quite long. Hence, the wave energy decreases only very gradually with height. As a consequence, the velocity amplitude increases with height because the decrease in density is steeper than the energy density (see middle panel of figure\,\ref{fig:7}).

It should be noted that the presence of a node in the line-forming layer has been observationally shown in some roAp stars (see Sect.\ref{sec:1} and \ref{sec:6.2}). Also it should be noted that, by taking account of magnetic effects on pulsation and an atmospheric model having a temperature inversion layer, \citet{Saio_etal10} demonstrated magneto-acoustic modes might have a node layer in the line-forming layer.

Figure\,\ref{fig:12} displays the line-profile variation of the H\,$\alpha$ line produced by an axisymmetric dipole p-mode at a sequence of the rotation phase $\phi=$0.0, 0.125, 0.25, 0.375 and 0.5 (from left to right). The rotational velocity at the equator is assumed to be $8.3\,{\rm km}\,{\rm s}^{-1}$. The rotation axis has been assumed to be inclined with respect to the line-of-sight by $90^\circ$, and the symmetry axis of pulsation has been assumed to be inclined with respect to the rotation axis of the star by $90^\circ$. 

Figure\,\ref{fig:13} displays the line-profile variation of Nd\emissiontype{III} $\lambda\, 6145.07$\,{\AA} line produced by the same mode and the same geometrical configuration as in the case of figure\,\ref{fig:12} at a sequence of the rotation phase $\phi=$0.0, 0.125, 0.25, 0.375 and 0.5 (from left to right). As seen in this figure, the residual is not symmetric at some rotational phases. This is caused by lateral inhomogeneity of the distribution of neodymium. Since we have assumed that the element is slightly concentrated in the magnetic polar region, which is misaligned with the rotation axis of the star, the line shape becomes significantly asymmetric when one magnetic polar region is coming to appear on the visible hemisphere while the other polar region is coming to disappear. The same effect can be seen in the H\,$\alpha$ line shown in figure\,\ref{fig:12}. This effect is most conspicuous in the case of $i=90^\circ$ and $\beta=90^\circ$.
\begin{figure*}[thb]
\begin{center}
\vspace{0.4cm}
\includegraphics[width=0.5\linewidth,angle=90]{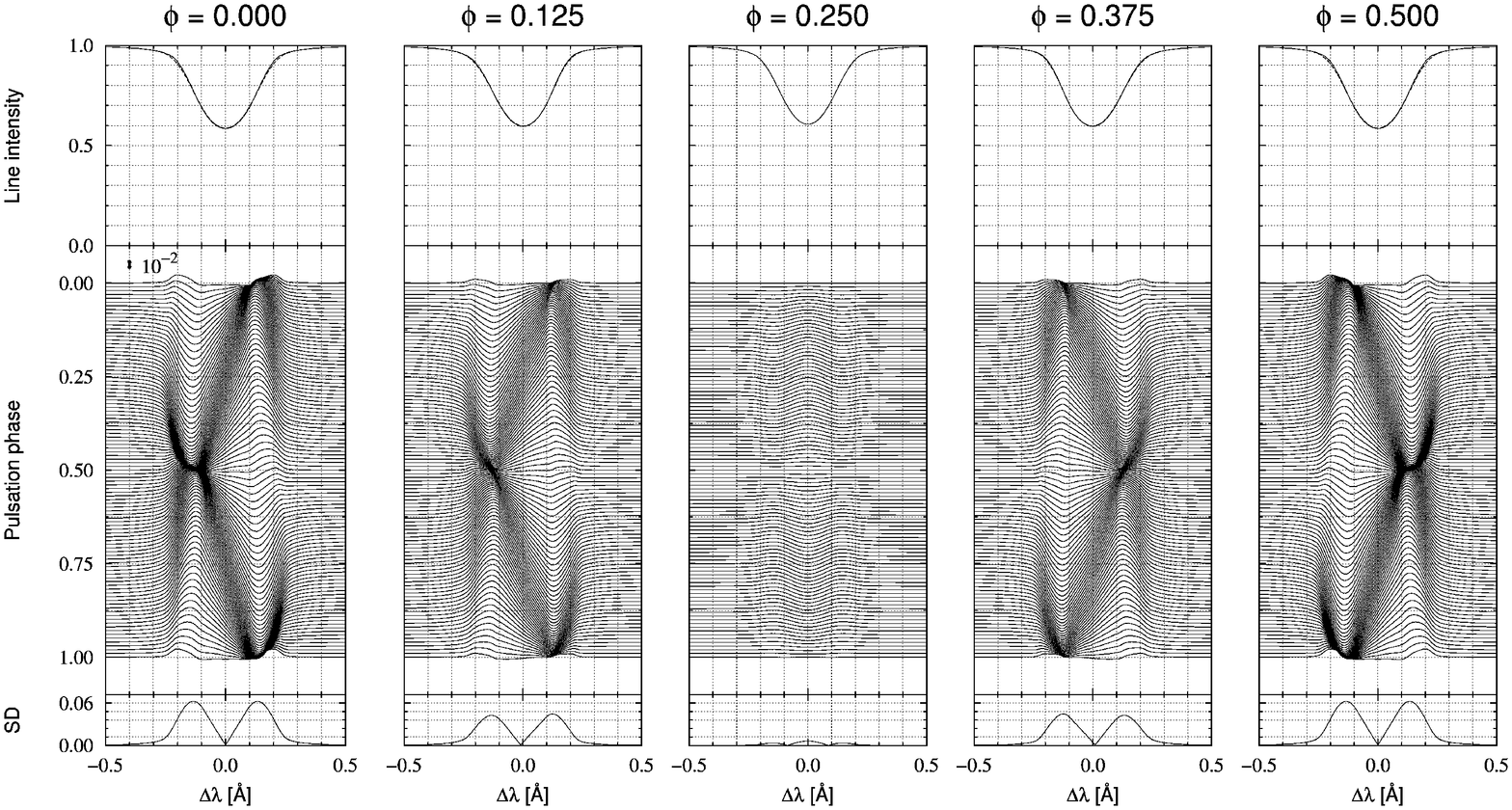}
\end{center}
\caption{Same as figure\,\ref{fig:12}, but for the Nd\emissiontype{III} $\lambda\, 6145.07$\,{\AA} line. }
\label{fig:13}
\end{figure*}

\subsubsection{Running wave}
\label{sec:5.4.2}
\noindent
Though the eigenmodes of the present model are evanescent above the photosphere and have no node there, spectroscopic observations of rapid oscillations of roAp stars indicate the presence of a nodal surface in the line-forming layers (e.g., \cite{Baldry_etal98}, \yearcite{Baldry_etal99}). Furthermore, in some stars, while H\,$\alpha$ line shows the standing wave feature, the absorption lines of Nd\emissiontype{III} and some other rare earth elements imply the running wave feature \citep{Kurtz_etal03}. Our present model is not fit to these Ap stars. The presence of a strong magnetic field and/or peculiar chemical composition in the atmosphere is likely to change substantially the atmospheric structure, and particularly the oscillation feature. In the case of a star with $B=10\,{\rm mT}$ $(= 1\,{\rm kG})$, the local Alfv\'en speed above the photosphere exceeds the sound speed. As a consequence, the acoustic wave and the magnetic wave couple with each other around the photospheric level, and the eigenmode in such a system is expected to be a mode with energy leakage (\cite{Sousa_Cunha08}; \cite{Shibahashi08}). We consider here an outwardly running wave and see its effect on the spectral line profile. 

We suppose that the wave energy is conserved during propagation through the line-forming layer. Then, the velocity amplitude grows with height in inverse proportion to the square root of density. Though, in the very outer layer, the amplitude may become supersonic, we assume here that the amplitude is still subsonic and sinusoidal, and estimate the vertical wavelength from the frequency and the sound speed. Figure\,\ref{fig:14} displays the line-profile variation of the H\,$\alpha$ line produced by such a running wave at a sequence of the rotation phase $\phi=$0.0, 0.125, 0.25, 0.375 and 0.5. The angle between the rotation axis of the star and the line-of-sight has been assumed to be $90^\circ$, and the symmetry axis of pulsation has been assumed to be inclined with respect to the rotation axis of the star by $90^\circ$. 
With propagation of the wave, the layer of the maximum amplitude shifts upward. Comparied with the line-profile variation in the case of a standing wave shown in figure\,\ref{fig:12}, we see that the running wave feature is characterized by the ridge structure migrating from blue to red and then back again to blue. 

The Nd\emissiontype{III} $\lambda\, 6145.07$\,{\AA} line forms at higher levels than the H\,$\alpha$ line, and so its line profile is deformed conspicuously around the phase at which the amplitude becomes large at the high atmosphere. This is seen as a spotty pattern shown in figure\,\ref{fig:15}, in which case we assume $i=90^\circ$ and $\beta=90^\circ$.   
\begin{figure*}[thb]
\begin{center}
\vspace{0.4cm}
\includegraphics[width=0.5\linewidth,angle=90]{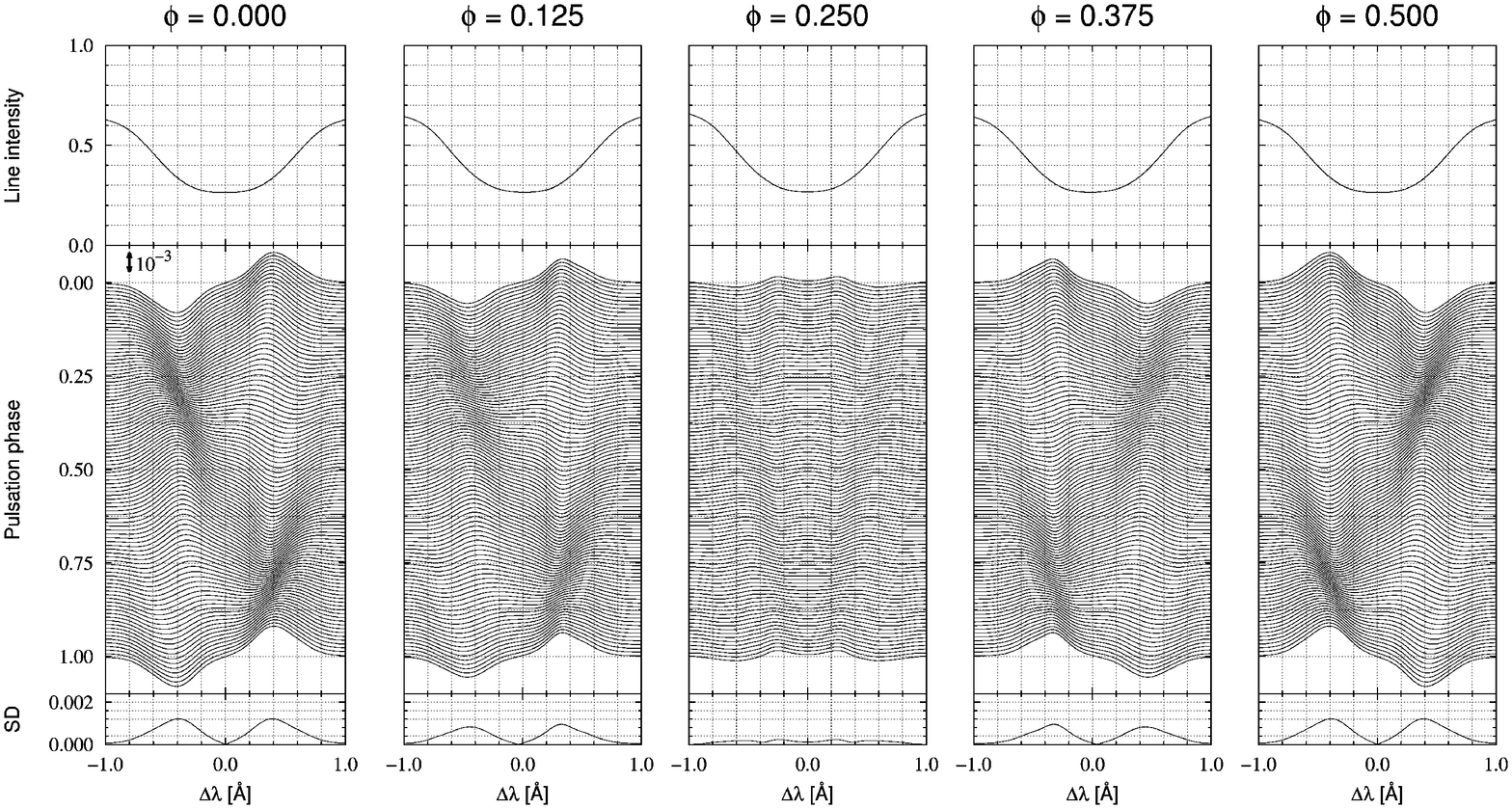}
\end{center}
\caption{Theoretically expected LPV of the H\,$\alpha$ line produced by an outward running wave of the axisymmetric dipole $(l=1, m=0)$ oscillation at a sequence of  the rotation phase $\phi=$ 0.0, 0.125, 0.25, 0.375, and 0.5 (from left to right). The rotation axis is assumed to be inclined with respect to the line-of-sight by $90^\circ$, and the symmetry axis of pulsation is assumed to be inclined with respect to the rotation axis of the star by $90^\circ$. 
In each panel, the differences between the individual profiles at a sequence of 100 pulsation phases and their mean are stacked with phase increasing downwards; above them is their mean, and below them is plotted their standard deviation. The abscissa is the wavelength measured from the line center in units of {\AA}. The left-hand ordinate scales refer to the phase of the oscillation, in units of the pulsation period.  
A short bar in the left panel indicates the length corresponding to $0.1\,\%$ of the continuum intensity.
}
\label{fig:14}
\end{figure*}
\begin{figure*}[bht]
\begin{center}
\vspace{0.4cm}
\includegraphics[width=0.5\linewidth,angle=90]{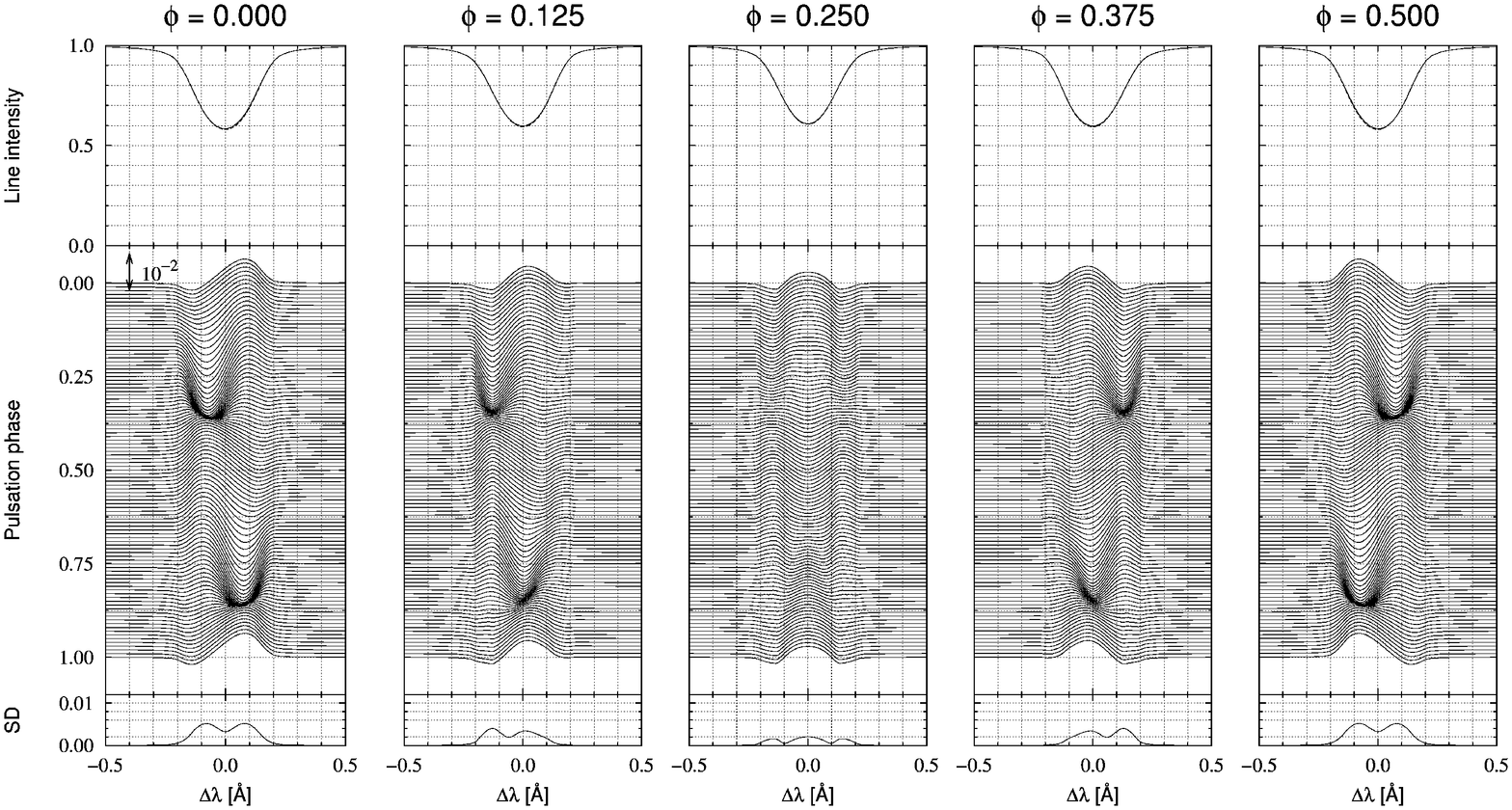}
\end{center}
\caption{Same as figure\,\ref{fig:14}, but for the Nd\emissiontype{III} $\lambda\, 6145.07$\,{\AA} line and the angle $i=90^\circ$ and $\beta=90^\circ$.}
\label{fig:15}
\end{figure*}

\section{Variation in bisectors}
\label{sec:6}
\subsection{Simulation results}
\label{sec:6.1}
\noindent
The variation in the bisector on the pulsation phase has been measured in some roAp stars. Let us consider the bisector motion expected from the present simulation. The most left panel of figure\,\ref{fig:16} shows the variation in the bisector of the H\,$\alpha$ line caused by the standing wave of axisymmetric, dipole, eigenmode on a sequence of pulsation phases in the case of $i=90^\circ$ and $\beta=90^\circ$. The amplitude of variation of the bisector gradually becomes larger with decrease in the residual intensity. The largest amplitude of variation in the bisector is apparently 
substantially smaller than the assumed largest velocity amplitude in the top of the atmosphere. This is partly because the density in the line-forming layer is higher than at the top of the atmosphere, and partly because  
the Doppler effect reflects only the line-of-sight velocity. 
As seen in the second left panel of figure\,\ref{fig:16}, the variation in the bisector in the case of a running wave is even smaller than in the case of a standing wave. This is because the standard deviation of the amplitude at a fixed height is smaller than that in the case of a standing wave.

\begin{figure*}[ht]
\begin{center}
\vspace{0.4cm}
\includegraphics[width=0.66\linewidth,angle=0]{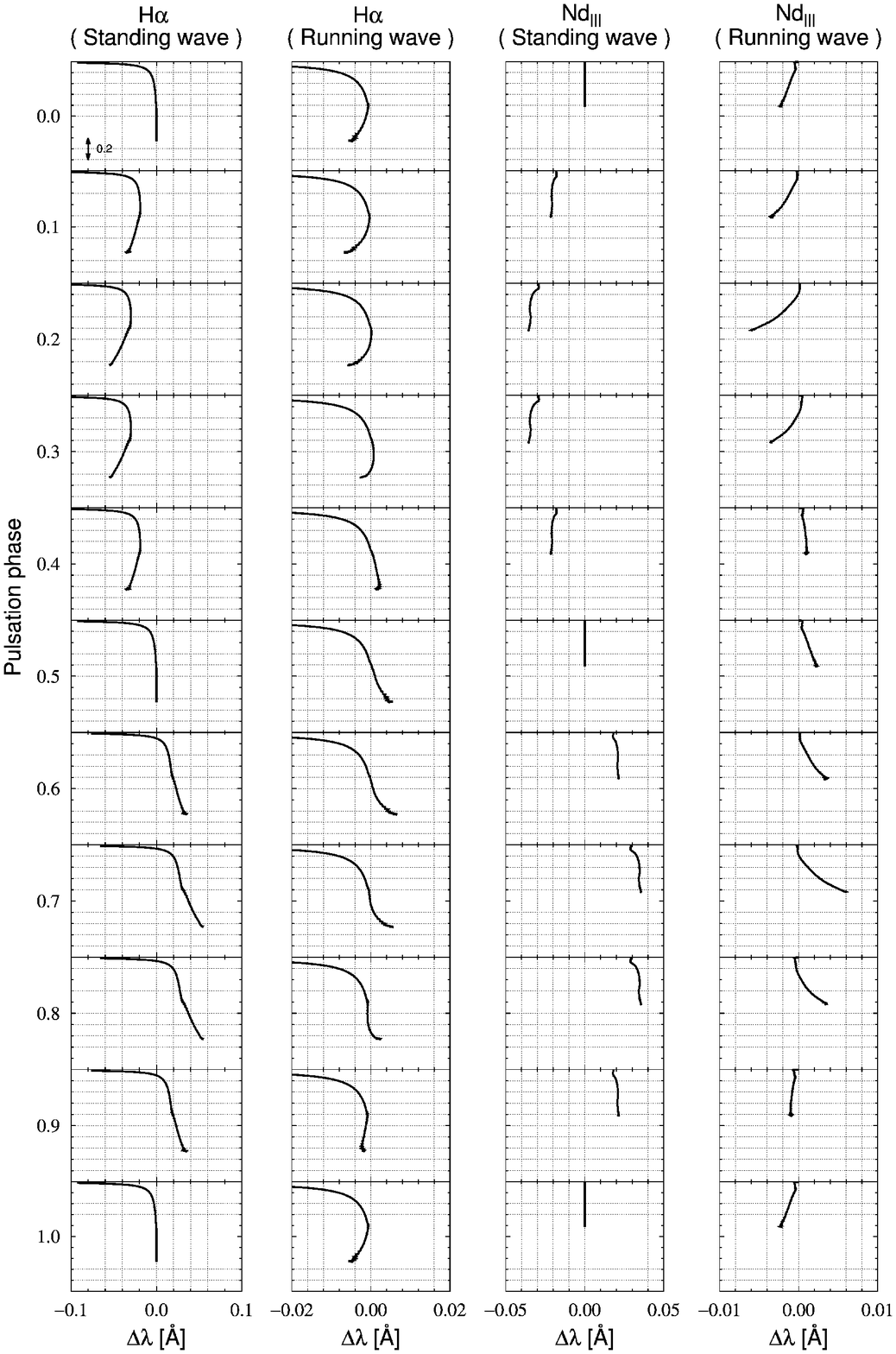}
\end{center}
\caption{
Theoretically expected variation in bisector of line profiles caused by axisymmteric dipole oscillation during a one complete period. The rotation axis and the pulsation axis are assumed to be aligned with the line-of-sight $(i=\beta=90^\circ)$. From left to right: H\,$\alpha$ in the case of a standing eigenmode; H\,$\alpha$ in the case of a running wave; Nd\emissiontype{III}\ $\lambda\,6145.07$\,{\AA} line in the case of a standing eigenmode; Nd\emissiontype{III}\ $\lambda\,6145.07$\,{\AA} line in the case of a running wave.  
In each panel, the line profiles (solid lines) and the bisectors (dashed lines) at the pulsation phases 0.0, 0.1, ..., 1.0 are stacked with phase increasing downwards. The abscissa is the deviation from the line center of the intrinsic profile, in units of {\AA}, and the ordinate of each diagram at a fixed pulsation phase is the line intensity normalized with the continuum level.  
A short bar in the left top panel indicates the length corresponding to $20\,\%$ of the continuum intensity.
}
\label{fig:16}
\end{figure*}
The variation in bisector of Nd\emissiontype{III} $\lambda\,6145.07$\,{\AA} line on the pulsation phase 
is shown in figure\,\ref{fig:16}. The second-right panel is the case of the standing wave of axisymmetric, dipole mode. The geometrical configuration is the same as in the case of the most-left panel. As seen in this panel, the general tendency is similar to the case of the H\,$\alpha$ line, though the range of variation is slightly smaller. 

The variation in the case of a running wave is displayed in the most-right panel. Compared with the case of  the H\,$\alpha$ line, the amplitude of variation becomes more conspicuous with a decrease in the residual intensity. This is because the line center of Nd\emissiontype{III}\ $\lambda\,6145.07$\,{\AA} line is formed around $\tau_{5000} \simeq 10^{-6.5}$, while the line center of H\,$\alpha$ line is sensitive to the layer around $\tau_{5000}\simeq 10^{-5.5}$. 

\subsection{Comparison with observations}
\label{sec:6.2}
\noindent
\citet{Mkrtichian_etal03} found in the roAp star HD\,137949 (33\,Lib) that the oscillation phases of Nd\emissiontype{III} lines are nearly $\pi$ out of phase with Nd\emissiontype{II} lines. This indicates presence of a nodal layer just beneath the Nd\emissiontype{III} line-forming layer of the star. They also suspected the presence of another nodal layer close to the continuum formation level.  
Similar results were obtained by \citet{Kurtz_etal05a}. They found that there is a phase jump between the Nd\emissiontype{II} line-forming layer and the Nd\emissiontype{III} line-forming layer, and concluded the presence of a nodal layer between these two types of lines. They also found that the amplitude of the Nd\emissiontype{III} line, which is formed in the higher layer than the Nd\emissiontype{II} line, is almost constant, although a small increase with height cannot be ruled out. 
The present calculation, however, does not show any node just beneath the Nd\emissiontype{III} line-forming layer, and this obviously means that fine tuning in making the equilibrium models and in mode calculation is necessary. Naively, an increase of the amplitude with height from the node is theoretically expected. 
Furthermore, \citet{Kurtz_etal05a} showed that the phase of the Nd\emissiontype{III} line is constant in the high atmosphere, but shows an outwardly running wave character in the deeper layer. This implies that the magnetic effect on the oscillations should be taken into account.  

\citet{Elkin_etal05a} found in the case of the roAp star HD\,99563 that the amplitude and phase of H\,$\alpha$ core vary strongly with the line depth. There seems a nodal layer within the H\,$\alpha$ line-forming layer. The same feature was found in some other rare erath lines including Nd\emissiontype{III} $\lambda\, 6145.07$\,{\AA} line. The present model does not show a node in the H\,$\alpha$ line-forming layer. This again means that fine tuning in making the equilibrium models and in mode calculation is necessary.

\citet{Ryabchikova_etal07a} analyzed the bisector motion of the H\,$\alpha$ and neodymium lines of some roAp stars. Qualitatively, the general feature of the bisector motion of H\,$\alpha$ line is in good agreement between the simulation and the observations, in the sense that the amplitude gradually becomes larger with the line depth. However, the situation is quite different in the case of the Nd\,III line. While the present simulation shows that the amplitude becomes larger with the line depth, the observations show the opposite tendency. This implies a possibility of strong dissipation in the high atmosphere, but more careful comparison should be made before concluding.

\citet{Kurtz_etal07} found an increase in amplitude of  the Nd\emissiontype{III} lines with line strength 
in the case of HD\,134214, and also found phase variations of those lines, suggesting an outwardly running wave. 
 
\citet{Elkin_etal08} studied LPV of the roAp star HD\,176232 (10\,Aql), and showed that the amplitude of the H\,$\alpha$ line increases with the atmospheric height, and that the phase variation on the atmospheric height indicates an outwardly running wave feature. They also found a similar tendency concerning the amplitude of Nd\emissiontype{III} lines to that found by \citet{Ryabchikova_etal07a}. That is, they found that the phase of the weak lines is almost constant, suggesting a standing wave, and that the amplitude decreases with the atmospheric height. 
Lines of Nd\emissiontype{III} with different intensities were found to have pulsation amplitudes different from the weaker lines that are formed more deeply in the atmosphere having higher amplitudes. It was also found that the bisectors for strong Nd\emissiontype{III} lines show significant changes of phase, indicating complex variations in the oscillation phase as a function of the atmospheric height.

\section{Discussion}
\label{sec:7}
\noindent
A simulation of LPV beyond single surface treatment is definitely necessary to obtain quantitative information of roAp stars from spectroscopic observations quoted in the previous section \ref{sec:6.2} with high time resolution, high spectral dispersion, and a high signal-to-noise ratio. This is a new information source of Ap stars, and is expected to be a promising new tool of asteroseismology to probe 3-D structure of atmospheres of these enigmatic stars.
Our aim is to establish an inversion approach of LPV to probe Ap stars' atmosphere and pulsation. 
This present paper is the very first step for this purpose, and there is much room for improvement. The following is a road map to our goal.
\begin{itemize}
\item
Levitation of chemical elements due to radiative pressure should be taken into account in model atmosphere.
\item
The Zeeman effect should be taken into account in calculating the line shapes of spectral lines.
\item
Nonadiabatic mode calculation should be done instead of adiabatic calculations.
\item
Magnetic effects on oscillations should be taken into account.
\item
The effect of temperature perturbation should be taken into account in LPV calculations.
\item
More systematic calculations should be carried out for many more lines.
\end{itemize}

\section*{Acknowledgments}
\noindent
We are very grateful to Masao Takata for helpful discussions.
A part of this work was supported by JSPS Grant-in-Aid for Scientific Research 23540260.


\end{document}